\documentclass[aps,prd,twocolumn,superscriptaddress,altaffilletter,lengthcheck,tightenlines,showpacs,showkeys]{revtex4-1}
\usepackage{color}
\usepackage{sidecap}
\usepackage[pdftex]{graphicx}
\usepackage{subfigure}
\usepackage{amsmath}
\newcommand{\be}{\begin{equation}}
\newcommand{\ee}{\end{equation}}
\newcommand{\n}{\label}
\newcommand{\no}{\noindent}
\newcommand{\ben}{\begin{eqnarray}}
\newcommand{\een}{\end{eqnarray}}
\usepackage{hyperref}

\begin{document}

\title{On extended sign-changeable interactions in the dark sector}

\author{M\'onica Forte}
\email{dramonicaforte@gmail.com}
\affiliation{Departamento de F\'isica, Facultad de ciencias Exactas y Naturales,
Universidad de Buenos Aires, 1428 Buenos Aires, Argentina}
\date{\today}

\begin{abstract}

We extend the cosmological couplings proposed in Sun et al. and Wei, where they suggested interactions with change of signs along the cosmological evolution. Our extension liberates the changes of sign of the interaction from the deceleration parameter and from the relation of energy densities of the dark sector and considers the presence of non interactive matter. In three cases we obtain the general solutions and the results obtained in models fitted with Hubble's function and SNe Ia data, are analyzed regarding the problem of the cosmological coincidence, the problem of the crisis of the cosmological age and the magnitude of the energy density of dark energy at early universe. Also we graphically study the range of variation of, the actual dark matter density parameter, the effective equation of state of the dark energy and  the redshift of transition to the accelerated regimen, generated by variations at order $1\sigma$ in the coupling parameters.
\end{abstract}
\pacs{}

\keywords{sign-changeable interactions, dark energy, dark matter, cosmological coincidence, crisis of age}

\maketitle

%%%%%%%%%%%%%%%%%%%%%%%%%%%%%%%%%%%%%%%%%%%%%%%%%%%%%%%%%%%%%%%%%%%%%%

\section{Introduction}
%%%%%%%%%%%%%%%%%%%%%%%%%%%%%%%%%%%%%%%%%%%%%%%%%%%%%%%%%%%%%%%%%%%%%

        Cosmological and astrophysical data from type Supernovae Ia (SNeIa) data, Cosmic Microwave Background (CMB) radiation anisotropies and Large Scale Structure (LSS), have provided strong evidences for a phase of accelerated expansion of our spatially flat universe \cite{Supernova,SDSS,Spergel:2003cb,Dark}. The dominant components in our current universe are dubbed dark energy (DE) and dark matter (DM) because of they share a non luminous nature.  They behave very differently, while DE has negative pressure and is responsible for the aforementioned acceleration, DM is gravitationally attractive and allows the accumulation of matter that leads to the formation of large scale structures. Only hypotheses on their nature exist, most of them assuming that DM and DE are physically unrelated and the similarity in their energy densities (the so called problem of Cosmological Coincidence) is a purely accidental fact \cite{Wetterich:1987fmyRatra:1987rm,Peebles:2002gy,Kolb:2005da}. But interactions between DM and DE might alleviate the problem, keeping close values for their densities up to large redshifts as has been shown in a lot of work \cite{Ellis_Wetterich_Amendola_Gasperini},\cite{interacc}, \cite{Chen:2011cy}. 
The interactions between DE and DM appear to be favored in several works, for instance, by Bertolami et al. \cite{Bertolamivarios} through studies on the Abell Cluster A586, and Abdalla et al. where they found the signature of interaction between DE and DM by using optical, X-ray and weak lensing data from 33 relaxed galaxy clusters \cite{Abdalla:2007rd}. Their conclusions are that this coupling is small but indicates that DE might be decaying into DM. Based on the detail analysis in perturbation equations of DE and DM when they are in interaction, He et al. find that the large scale CMB, especially the late Integrated Sachs Wolfe effect, is a useful tool to measure the coupling between dark sectors and arrive at the same conclusion \cite{He:2009pd}. 
Although the observational constraints indicate the possibility that the decay can be in both directions, \cite{Guo:2007zk,Quartin:2008px}, the decay of DE into DM seems to be strongly favored from a thermodynamical perspective, provided that the chemical potential of both components vanish or even without considering the chemical potential,\cite{Pavon:2007gt}. Other authors argue that If the chemical potential of at least one of the fluids is not zero, the decay can occur from the DM to DE, with no violation of the second law \cite{Pereira:2008at,Lima:2008uk,Pereira:2008af,Pereira:2008nd}.  
Also, in \cite{Olivares:2008bx} a restriction on the interaction parameter strength are obtained, discarding interacting models  with constant coupling strength greater than 0.1 at $2\sigma$ confidence level. This constraint has been obtained for the particular case of an interaction in the dark sector, which is proportional to the total energy density when other components are not considered.
Without a firm theoretical basis to raise the type of coupling, interactions are proposed and analyzed for compatibility with a number of basic premises arising from the huge amount of observational data. 
The cosmological coincidence problem (CC), the bounds on the dark energy density parameter in the early universe (EDE), the magnitude of the EoS of dark energy in the present and in the recent past, the possibility or necessity of nonzero pressure of dark matter, the problem of having milestones older than the age of the universe, the existence and variability of the transition redshift, are all matters to meet with the interaction under study. 
In addition, recently one has found that the interaction can change its sign, in contemporary times with that of the transition to the accelerated regime and not too distant to it, cosmologically speaking, but not necessarily coincidental with it \cite{Cai:2009ht}.
In light of these considerations, among the various possible  interactive proposals,  Cai's work indicates that should be selected those which can change its sign along cosmological evolution. There are a few papers \cite{Changeable}, where the proposals are explicit but analytical solutions are obtained only in some cases.
In this paper we propose an extension of these works, considering interactions where the deceleration parameter and the interaction do not change their signs together but with a short time delay and also, a baryonic component that is not coupled to dark sector is added. The selected sign-changeable interactions used here affect only the dark sector and in all proposed cases, our work  generates exact solutions. We analyze their behaviors through models whose parameters are adjusted minimizing a $\chi^2$ function  \cite{Press} with the Hubble data $H(z)$ \cite{Stern:2009ep,Simon:2004tf,Moresco:2012by,Farooq:2013hq,Liao:2012bg}, and SNIa observations  \cite{Riess:2009pu}. The H(z) test was probably first used to constraint cosmological parameters in \cite{Samushia:2006fx} and then in a large number of papers \cite{Wei:2006ut,Lazkoz:2007zk,Lin:2008wr,Cao:2011cg,Figueroa:2008py,Seikel:2012cs,Santos:2011cj,delCampo:2010zz,Aviles:2012ir}. In the next section we present the general framework which describes the dark sector with an exchange of energy and the evolution of the non interacting fluid. After that, in section III, we analyze explicit interactions with worked examples based on models with adjusted parameters. In section IV we compare the results obtained and in the last section, we summarize our main outcomes and conclude.

%%%%%%%%%%%%%%%%%%%%%%%%%%%%%%%%%%%%%%%%%%%%%%%%%%%%%%%

\section{The models}

%%%%%%%%%%%%%%%%

We consider cosmological models with two interacting dark fluids plus a non interacting baryonic component in flat Friedmann Robertson Walker metric (FRW). The dark sector contains a dark matter fluid with energy density $\rho_1$ in interaction with a dark energy fluid with energy density $\rho_2$. The scenario is completed with a not interactive matter component with energy density $\rho_m$ that at first, can be dust or radiation. We have assumed that the equations of state (EoS) are $p_i=(\gamma_i-1)\rho_i$ for constant barotropic indexes $\gamma_i$, $i = 1,2,m$. Also, we have defined an auxiliary "dark" barotropic index $\gamma=(\gamma_1\rho_1+ \gamma_2\rho_2)/(\rho_1+\rho_2)$ and the overall effective barotropic index  $\gamma_t=(\gamma\rho+ \gamma_m\rho_m)/(\rho+\rho_m)$. The Friedmann equation and the conservation equation are given by
\begin{subequations}
\n{01}
\be
\n{01a}
\textstyle{3H^2=\rho_t=\rho+\rho_m }
\ee
\vspace{-0.9cm}
\be
\n{01b}
\textstyle{\rho=\rho_1+\rho_2},
\ee
\end{subequations}
\vspace{-0.6cm}
\begin{subequations}
\n{02}
\be
\n{02a}
\textstyle{\rho_t'+\gamma_t\rho_t=0}
\ee
\vspace{-0.9cm}
\be
\n{02b}
\textstyle{\rho'+\gamma\rho=0}
\ee
\vspace{-0.9cm}
\be
\n{02c}
\textstyle{\rho_m'+\gamma_m\rho_m=0}.
\ee
\end{subequations}

Above and henceforth $'$ means derivative with respect to the variable $\eta = \ln(a/a_0)^3$ and $a$, $a_0$ and $H=\dot a / a$ are the factor of scale for the FRW metric, the actual factor of scale and the Hubble parameter respectively. Using (\ref{01b}), (\ref{02b})  and (\ref{02c}) the partial energy densities $\rho_1$, $\rho_2$ and $\rho_m$ can be written as

\begin{subequations}
\n{03}
\be
\n{03a}
\textstyle{\rho_1= \rho\frac{\gamma-\gamma_2}{\gamma_1-\gamma_2}},
\ee
\vspace{-0.5cm}
\be
\n{03b}
\textstyle{\rho_2= \rho\frac{\gamma_1-\gamma}{\gamma_1-\gamma_2}},
\ee
\vspace{-0.5cm}
\be
\n{03c}
\textstyle{\rho_m= \rho_{0m}(1+z)^{3\gamma_m}}, \qquad 1+z = a_0/a.
\ee
\end{subequations}

\vskip.2cm
The interaction $\mathcal{Q}$, that is  only applied to the dark sector, is defined through the splitting of the equation of conservation (\ref{02b}) as
\begin{subequations}
\n{04}
\be
\n{04a}
\textstyle{\rho_1'+ \gamma_1 \rho_1= -\mathcal{Q}},
\ee
\be
\n{04b}
\textstyle{\rho_2'+ \gamma_2 \rho_2= \mathcal{Q}}.
\ee
\end{subequations}
\vskip.2cm

Once the interaction $\mathcal{Q}$ is fixed, the dynamic evolution of the auxiliar dark EoS $\gamma$ is obtained solving the integro-differential equation

\be
\n{05}
\textstyle{\gamma' - (\gamma - \gamma_1)(\gamma - \gamma_2)= - \mathcal{Q} (\gamma_1 - \gamma_2) e^{\int \gamma d\eta}}.
\ee
\vskip.2cm

Sometimes, the general solution of (\ref{05}), $\gamma(\eta) = - d\ln \rho/d\eta$, can be integrated to obtain the expression of the energy density of the dark sector $\rho$. In such cases, we have all the necessary information to know the partial dark densities through the equations (\ref{03}), the dark ratio $r(z)=\rho_{DM}(z)/\rho_2(z)$, the effective EoS for the dark sector $\omega(z)=\gamma(z) - 1$, the deceleration parameter $ q (z) = - 1 +3 \gamma_t (z)/2 $ and the effective partial state equations $\omega_{eff1}(z)=\omega_{1}+\mathcal{Q}(\gamma,\rho)/\rho_1$ and $\omega_{eff2}(z)=\omega_{2}-\mathcal{Q}(\gamma,\rho)/\rho_2$.
 With that knowledge, we obtain the explicit expression of $ \mathcal{Q}(z) $, and as well, the benefits or defects of each interaction can be calibrated.   Among other things, we can analyze if the density parameter of dark energy at early times (EDE) respects the constraints arising from the baryogenesis; if the redshift of sign change $z_{\mathcal{Q}}$ is consistent with the results of Cai et al., and if the magnitude of the ratio r in the final stage of the universe alleviates the coincidence problem and to what extent it does.
A variety of interactions, which can change its sign, has been proposed \cite{Changeable}. The Wei's work matches $z_{\mathcal{Q}}$ with the transition redshift to an accelerated universe $z_{acc}$, while Sun et al. linking  $z_{\mathcal{Q}}$ with a redshift for which the dark ratio r equals the inverse of the coupling constant (really there, only the case $r = 1$ is considered). These studies limit themselves to the dark sector.
In the following sections, we extend these studies proposing models with interactions in the dark sector, that show no explicit connection between both redshifts, $z_{\mathcal{Q}}$ and $z_{acc}$ plus a fluid satisfying its own  conservation equation.

\section{ SIGN-CHANGEABLE INTERACTIONs}

Literature does not provide a natural guidance from fundamental physics on the cosmological interaction $\mathcal{Q}$ and so we can only discuss it to a phenomenological level.  The most familiar cosmological interactions used are proportional to the energy densities $\rho_{de}$, $\rho_{dm}$ and $\rho$, and these magnitudes always can be written in terms of the energy density $\rho$ and of its derivative $\rho'$. So, it seems natural that interactive term corresponds to linear combinations $L(\rho,\rho')=d_1\rho+d_2\rho'$ of this type or more generally, that $\mathcal{Q}$ be a non lineal function $G(\rho,\rho')$. In general the models with linear combinations  do not allow realizing a change in the direction of transfer of energy between DE and DM and we do not know a solid theory which provides a Lagrangian to derive the appropriate interaction term. Therefore, we propose phenomenological multiplicative terms including the deceleration parameter $q$ plus an statistically adjustable degree of freedom $\sigma$ that allows differentiate the transition deceleration - acceleration of the universe of the directional transition on the transfer of energy. That is,  we propose non linear functions $G(\rho,\rho')=(\sigma -1+3\rho'/2\rho)L(\rho,\rho')$.

\subsection{$3H\mathcal{Q}_1 = (\sigma-1+\frac{3}{2}\gamma)(\alpha \dot\rho_2+3H\beta\rho_2)$}

The sign-changeable interaction 
\be
\n{06}
\textstyle{\mathcal{Q}_1 = (\sigma-1+\frac{3}{2}\gamma)(\alpha \rho_2'+\beta\rho_2)},
\ee
is a natural extension of that proposed by Wei (and solved for $\alpha = 0$), which is reobtained when $\sigma$ and $\rho_m$ are null. It can be written in terms of the auxiliary "dark" barotropic index $ \gamma$, using (\ref{02b}) and (\ref{03b}) as 
\be
\n{07}
\textstyle{-\frac{(\gamma_1 - \gamma_2)\mathcal{Q}_1}{\rho}}= \textstyle{(\sigma-1+\frac{3}{2}\gamma)\Big(\alpha \gamma'+(\gamma_1 - \gamma)(\alpha \gamma - \beta)\Big)}.
\ee
 For this option, the equation (\ref{05}) results into two different first order differential equations for $\gamma$ depending on, whether or not $\alpha$ is zero:

\subsubsection{\bf{$\alpha=0$}}

In the case $\alpha=0$, the interaction (\ref{07}) leads to the first order differential equation for $\gamma$
\be
\n{08}
\textstyle{\gamma'+ (1+\frac{3}{2}\beta)(\gamma_1-\gamma)\Big(\gamma - \frac{(\gamma_2+\beta-\sigma\beta)}{1+\frac{3}{2}\beta}\Big)=0},
\ee
whose general solution is
\be
\n{09}
\textstyle{\gamma(\eta)}=\textstyle{\frac{(1-c_1)\gamma_0 + c_1\gamma_1e^{-\eta u (\gamma_1-\gamma_0)}}{(1-c_1)+ c_1 e^{-\eta u (\gamma_1-\gamma_0)}}},
\ee
\no  with $u=(1+\frac{3}{2}\beta)$  and $\gamma_0 = (\gamma_2+\beta - \sigma\beta)/u$.
From (\ref{02b}) and (\ref{09}) we obtain the general solution for $\rho(z)$
\be
\n{10}
\textstyle{\frac{\rho(z)}{\bar\rho_0}}=\textstyle{\Big(c_1 (1+z)^{3u\gamma_1}+(1-c_1)(1+z)^{3u\gamma_0}\Big)^{\frac{1}{u}},}
\ee
\no where $\bar\rho_0$ and $c_1$ are constants of integration.\\

The solutions (\ref{09}) and (\ref{10}) allow to obtain, through the expressions (\ref{01a}) and (\ref{03}), the functional forms of the total energy density $\rho_t$, the partial energy densities $\rho_1$ and $\rho_2$, and the dark ratio $r=\rho_1/\rho_2$,
\be
\n{011}
\textstyle{\frac{\rho^{(\mathcal{Q}_{1_{\alpha=0}})}_t(z)}{\rho_0}}=\textstyle{b_mx^{\gamma_m}+(1-b_m)\Big(c_1 x^{u\gamma_1}+(1-c_1) x^{u\gamma_0}\Big)^{\frac{1}{u}}}, 
\ee

\ben
\begin{aligned}
\n{012}
\rho^{(\mathcal{Q}_{1_{\alpha=0}})}_1&(z)= \textstyle{\frac{\bar\rho_0\Big(c_1 x^{u\gamma_1}+(1-c_1) x^{u\gamma_0}\Big)^{\frac{1-u}{u}}}{(\gamma_1-\gamma_2)}}\\
&\textstyle{\times \Big(c_1(\gamma_1-\gamma_2)x^{u\gamma_1}+(1-c_1)(\gamma_0-\gamma_2)x^{u\gamma_0}\Big)},\ \ 
\end{aligned}
\een

\ben
\n{013}
\textstyle{\frac{\rho^{(\mathcal{Q}_{1_{\alpha=0}})}_2(z)}{\bar\rho_0(\gamma_1-\gamma_0)}}= \textstyle{\frac{(1-c_1)x^{u\gamma_0}\Big(c_1 x^{u\gamma_1}+(1-c_1) x^{u\gamma_0}\Big)^{\frac{1-u}{u}}}{(\gamma_1-\gamma_2)}},
\een
\no and 

\ben
\begin{aligned}
\n{014}
\textstyle{r^{(\mathcal{Q}_{1_{\alpha=0}})}}\textstyle{(z)=\frac{\gamma_0-\gamma_2}{\gamma_1-\gamma_0}+\frac{c_1(\gamma_1-\gamma_2)x^{u(\gamma_1-\gamma_0)}}{(1-c_1)(\gamma_1-\gamma_0)}}
\end{aligned}
\een
\vskip0.5cm
\no where\ \  $x=(1+z)^3$,\ \  $b_m=\rho_{0m}/\rho_0$,\ \  $\bar\rho_0=\rho_0(1-b_m)$  and $\rho_0$ is the actual total energy density. 

The explicit expressions for the interaction $\mathcal{Q}_{1_{\alpha=0}}$, the deceleration parameter $q$, the density parameter for the dark energy $\Omega_2$, and the effective equations of state  \ \  $\omega_{1eff}$, $\omega_{2eff}$ and $\omega_{t}$, are
\be
\begin{aligned}
\n{015}
&\textstyle{\mathcal{Q}_{1_{\alpha=0}}}(z)= \textstyle{\rho_0\frac{(\gamma_1-\gamma_0)}{(\gamma_1-\gamma_2)}\Big(c_1 x^{u\gamma_1}+(1-c_1) x^{u\gamma_0}\Big)^{\frac{1-2u}{u}}\times} \\ 
& \textstyle{\beta(1-b_m)(1-c_1)x^{u\gamma_0}\Big(c_1\frac{\nu_1}{2} x^{u\gamma_1}+(1-c_1)\frac{\nu_0}{2} x^{u\gamma_0}\Big)},\ \   
\end{aligned}
\ee

\be
\begin{aligned}
\n{016}
&q^{(\mathcal{Q}_{1_{\alpha=0}})}(z)=\textstyle{\frac{b_m(\frac{3}{2}\gamma_m-1)x^{\gamma_m}}{b_mx^{\gamma_m}+(1-b_m)\Big(c_1 x^{u\gamma_1}+(1-c_1) x^{u\gamma_0}\Big)^{\frac{1}{u}}}+}\scriptsize{(1-b_m)} \\
& \textstyle{\times\frac{\Big(c_1 x^{u\gamma_1}+(1-c_1) x^{u\gamma_0}\Big)^{\frac{1-u}{u}}\Bigg(c_1 (\frac{3}{2}\gamma_1-1)x^{u\gamma_1}+(1-c_1)(\frac{3}{2}\gamma_0-1) x^{u\gamma_0}\Bigg)}{b_mx^{\gamma_m}+(1-b_m)\Big(c_1 x^{u\gamma_1}+(1-c_1) x^{u\gamma_0}\Big)^{\frac{1}{u}}}},\ \ \ 
\end{aligned}
\ee

\be
\n{017}
\Omega^{(\mathcal{Q}_{1_{\alpha=0}})}_2(z)=\textstyle{\frac{(1-b_m)(1-c_1)(\gamma_1-\gamma_0) x^{u\gamma_0}\Big(c_1 x^{u\gamma_1}+(1-c_1) x^{u\gamma_0}\Big)^{\frac{1-u}{u}}}{(\gamma_1-\gamma_2)\Big(b_mx^{\gamma_m}+(1-b_m)\big(c_1 x^{u\gamma_1}+(1-c_1) x^{u\gamma_0}\big)^{\frac{1}{u}}\Big)},}
\ee

\be
\begin{aligned}
\n{018}
\omega&^{(\mathcal{Q}_{1_{\alpha=0}})}_{1eff}(z)=\textstyle{\omega_{1}+\frac{\beta(1-c_1)(\gamma_1-\gamma_0) x^{u\gamma_0}}{\Big(c_1 x^{u\gamma_1}+(1-c_1) x^{u\gamma_0}\Big)}\times} \\
&\textstyle{\frac{\Big(c_1 \frac{\nu_1}{2} x^{u\gamma_1}+(1-c_1)\frac{\nu_0}{2}  x^{u\gamma_0}\Big)}{\Big(c_1(\gamma_1-\gamma_2) x^{u\gamma_1}+(1-c_1)(\gamma_0-\gamma_2) x^{u\gamma_0}\Big)},}
\end{aligned}
\ee
\be
\n{019}
\omega^{(\mathcal{Q}_{1_{\alpha=0}})}_{2eff}(z)=\textstyle{\omega_{2}-\frac{\beta\Big(c_1\frac{\nu_1}{2} x^{u\gamma_1}+(1-c_1)\frac{\nu_0}{2}  x^{u\gamma_0}\Big)}{(c_1x^{u\gamma_1}+(1-c_1) x^{u\gamma_0})},}
\ee

\be
\begin{aligned}
\n{020}
&\omega^{(\mathcal{Q}_{1_{\alpha=0}})}_t(z)=\textstyle{\frac{b_m \omega_m x^{\gamma_m}}{b_m x^{\gamma_m}+(1-b_m)\Big(c_1 x^{u\gamma_1}+(1-c_1) x^{u\gamma_0}\Big)^{\frac{1}{u}}}} + \\
&\textstyle{\frac{(1-b_m)\Big(c_1 x^{u\gamma_1}+(1-c_1) x^{u\gamma_0}\Big)^{\frac{1-u}{u}}\Big[c_1\omega_1 x^{u\gamma_1}+(1-c_1)(\gamma_0-1)x^{u\gamma_0}\Big]}{b_m x^{\gamma_m}+(1-b_m)\Big(c_1 x^{u\gamma_1}+(1-c_1) x^{u\gamma_0}\Big)^{\frac{1}{u}}}.}
\end{aligned}
\ee
\vskip0.3cm
\no where $\nu_i \equiv 3\gamma_i + 2(\sigma-1)$, $i=0,1$.
\vskip0.3cm

          Here and in the following cases we provide a better approach to the interaction, studying its behavior with models whose parameters have been adjusted  using the Hubble data as in \cite{Stern:2009ep,Simon:2004tf,Press}.
The statistical analysis is based on the minimization of a $\chi^2$ function with the Hubble data which is constructed as 
\be
\n{021}
\chi^2(\Theta_{\mathcal{Q}_i})= \sum_{i=1}^{29} \frac{[H_{th}(\Theta_{\mathcal{Q}_i};z_i) - H_{obs}(z_i)]^2}{\sigma^2(z_i)}
\ee

\no where $\Theta_{\mathcal{Q}_i}$ stands for cosmological parameters involved in the specific interaction ${\mathcal{Q}_i}$, $H_{obs}(z_k)$ is the observational H(z) data at the redshift $z_k$, $\sigma^2(z_k)$ is the corresponding $1\sigma$ uncertainty, and the summation is over the 29 observational H(z) data \cite {Liao:2012bg}. In this case, where $\Theta_{\mathcal{Q}_{1_{\alpha=0}}}=(H_0,b_m,c_1,\gamma_m,\gamma_1,\gamma_2,\beta,\sigma)$ and $H_{th}(\Theta_{\mathcal{Q}_{1_{\alpha=0}}}; z_i)= \sqrt{\rho_t(z_i)/3}$ is taken from (\ref{011}),  we find $\chi^2_{min}=17.1474$, corresponding to  $\chi^2_{dof}=0.816543$, for the best fit values: $H_0=71.81~{\rm km s^{-1}\,Mpc^{-1}}$, $b_m=0.225$, $c_1=0.05$, $\gamma_m=1.00$, $\gamma_1=1.00$, $\gamma_2= 10^{-7}$, $\beta=-0.10$ and $\sigma=0.60$. \\
Then, the best adjusted model for $\mathcal{Q}_{1_{\alpha=0}}$, exemplifies a universe with interaction between cold dark matter (CDM) and cosmological constant $\Lambda $ in presence of non-interacting dust. See Table \ref{tab:tabla1}.

%%Alpha0Fig1
\begin{figure}
\centering 
\begin{minipage}[t][9cm][b]{0,5\textwidth}
\includegraphics[height=20cm,width=18cm]{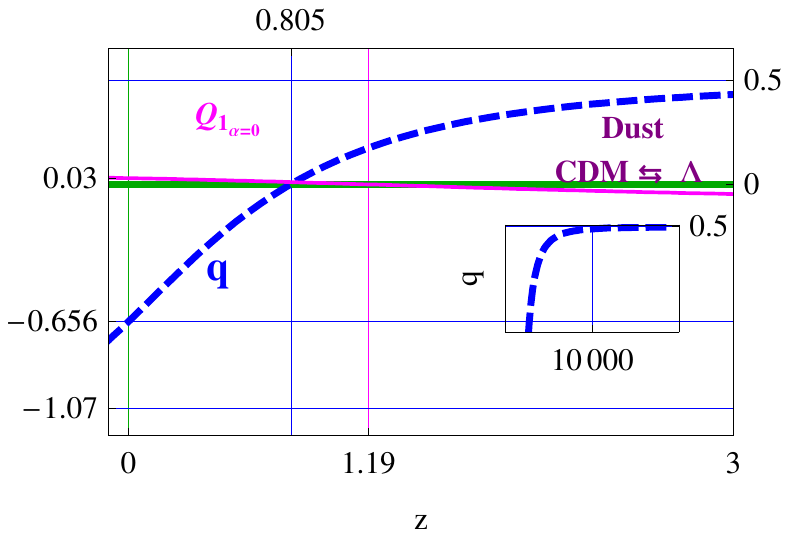}
\caption{\scriptsize{Evolution of the interaction $\mathcal{Q}_{1_{\alpha=0}} = \beta(\sigma-1+\frac{3}{2}\gamma)\rho_2$ (solid magenta curve) and corresponding deceleration parameter $q $ (dashed blue curve) for the best fit model  with parameters $H_0=71.81~{\rm km~s^{-1}\,Mpc^{-1}}$ , $b_m=0.225$, $\gamma_1=\gamma_m=1$, $\gamma_2=0$, $\beta=-0.1$, $\sigma=0.6$ and $c_1= 0.05$. With a  minimum $\chi^2_{dof}=0.816543$ per degree of freedom, this interactive cosmological model supports a coupling between CDM and cosmological constant $\Lambda $ in the presence of non interactive dust. $q $ tends to 0.5 at early times and to -1.07  in the distant future. Transition to the accelerated regime is checked at $z_{acc} = 0.8$, different of the  redshift at which the interaction changes its sign $z_{\mathcal{Q}_{1_{\alpha=0}} } = 1.19$ from which energy is transferred from the DM to the DE.}}
\label{Fig:.Alpha0Fig1}
\end{minipage}
\end{figure}

\begin{figure}
\centering 
\begin{minipage}[t][8cm][b]{0,5\textwidth}
\includegraphics[height=18cm,width=15cm]{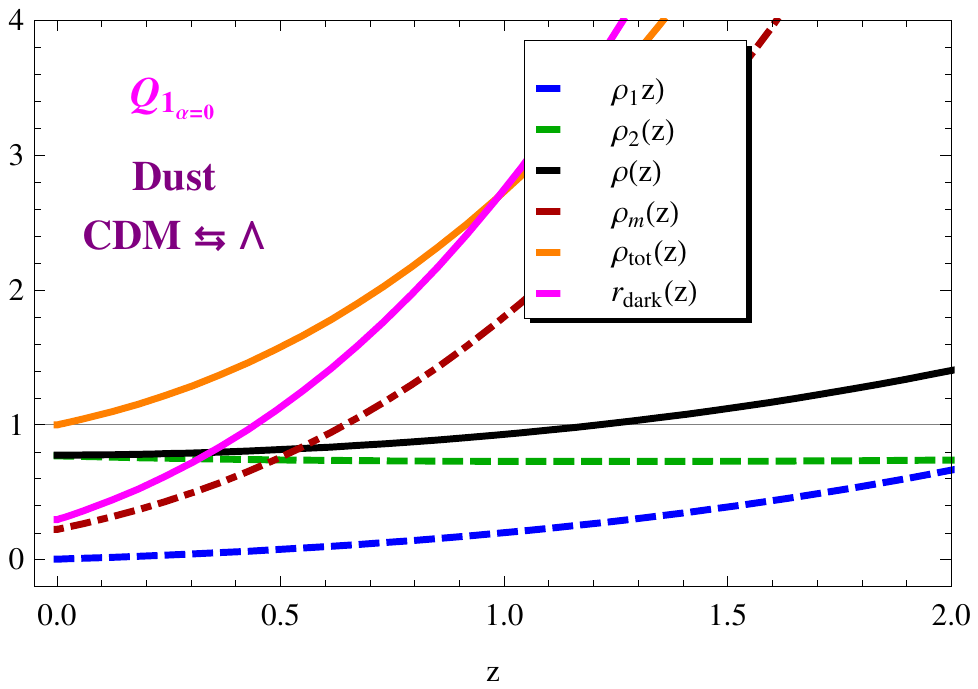}
\caption{\scriptsize{Energy densities and ratio of dark  densities for the best fit model with the interaction $\mathcal{Q}_{1_{\alpha=0}}$. The contribution to the total energy density $\rho_{t}$ comes mainly from non interactive matter (mostly dark matter) until $z \sim 0.5$ and the ratio of dark fluids becomes 1 at $z^{\scriptsize{(\mathcal{Q}_{1_{\alpha=0}})}} = 0.443$. The dark matter fluid used to calculate this ratio, in this case, consists of dark matter that participates in the interaction but also by the  part of the conserved fluid which cannot be considered baryonic as suggested by its density parameters. See Table \ref{tab:tabla2}.}}
\label{Fig:.Alpha0Fig2}
\end{minipage}
\end{figure}

\begin{figure}
\centering 
\begin{minipage}[t][9cm][b]{0,4\textwidth}
\includegraphics[height=25cm,width=18cm]{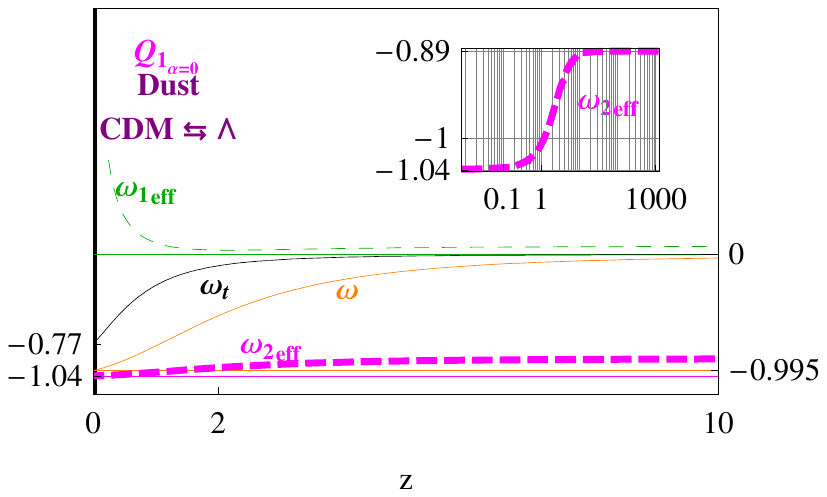}
\caption{\scriptsize{Evolution of the all effective equations of state involved in the best fit model for the interaction $\mathcal{Q}_{1_{\alpha=0}}$. At early times, $\omega_{t}$  and $\omega_{1eff}$,  both tend to zero, showing the dominance of dust fluid. The dark energy fluid has an asymptotic value  $\omega_{2eff}=-0.89$ at early times and crosses the PDL just before the present time where has the value $\omega_{2eff}=-1.04$.}}
\label{Fig:.Alpha0Fig3}
\vskip -0.5cm
\end{minipage}
\end{figure}

In Fig.\ref{Fig:.Alpha0Fig1} we show how the interaction $\mathcal{Q}_{1_{\alpha=0}}$ progresses from negative values in the past, up to the actual value $0.09 H_0^2$ changing its sign at $z_{\mathcal{Q}_{1_{\alpha=0}}} = 1.19$. Also, we see that the deceleration parameter has asymptotic values $0.5$ at early times and $ - 1.07$ in the far future, transits from a decelerated universe to an accelerating one at $z_{acc}=0.8 $  while its actual value is $q_0^{({\alpha=0})}= -0.656$. All these values are very much like those of the $\Lambda$CDM model, perhaps because most  non interactive dust, is dark, as it can be seen in Fig.\ref{Fig:.Alpha0Fig2} and in Table \ref{tab:tabla2}.  However, the interaction effect is observed as a delay in transition redshift (about $12.5\%$) respect to a $\Lambda$CDM with the same material density parameter $ b_m = 0.225 =\Omega_m(0) $. The strength of the interaction in this model is consistent with the result obtained in \cite{Olivares:2008bx} which sets an upper bound ($ < 0.1 $ at $ 99.95\%$ CL) through the discrepancy between the fraction of DE necessary to explain the amplitude of the ISW effect  larger than the allowed by the WMAP data. 
\vskip 2.5cm
%%Alpha0Fig2
\vskip 0.5cm
The Fig.\ref{Fig:.Alpha0Fig2} shows the evolution  of all energy densities involved in the model and the dark ratio between DM (interactive and non interactive) and DE fluids, that takes the value 1 around $z \sim 0.443$. 
The evolutions of all effective equations of state are shown in Fig.\ref{Fig:.Alpha0Fig3}, where we can see that at early times, $\omega_{t}$  and $\omega_{1eff}$,  both tend to zero, showing the absolute dominance of dust fluid. The inset shows that the dark energy fluid has an asymptotic value  $\omega_{2eff}=-0.89$ at early times and crosses the phantom divide line (PDL)  just before the present time where has the value $ \omega_{2eff}=-1.04$. We are interested in knowing how the magnitudes of the energy density parameter \small{$\Omega_{M0}\equiv \Omega_1(0)+\Omega_m(0)$}, the effective EoS  of dark energy  $\omega_{2eff}$ and the redshift of the transition to the accelerated regimen $z_{acc}$ are affected because of the variations in the coupling parameters $\beta$ and $\sigma$. For this task we take the $1\sigma$ CL intervals $\beta \ \epsilon\ \  [-0.37,0.61]$ and $ \sigma\ \   \epsilon\ \   [-3.9,7.3]$, (see Table \ref{tab:tabla2}).\\

%Alpha0Fig3

\begin{figure}
\centering 
\begin{minipage}[t][12cm][b]{0,5\textwidth}
\includegraphics[height=15cm,width=10cm]{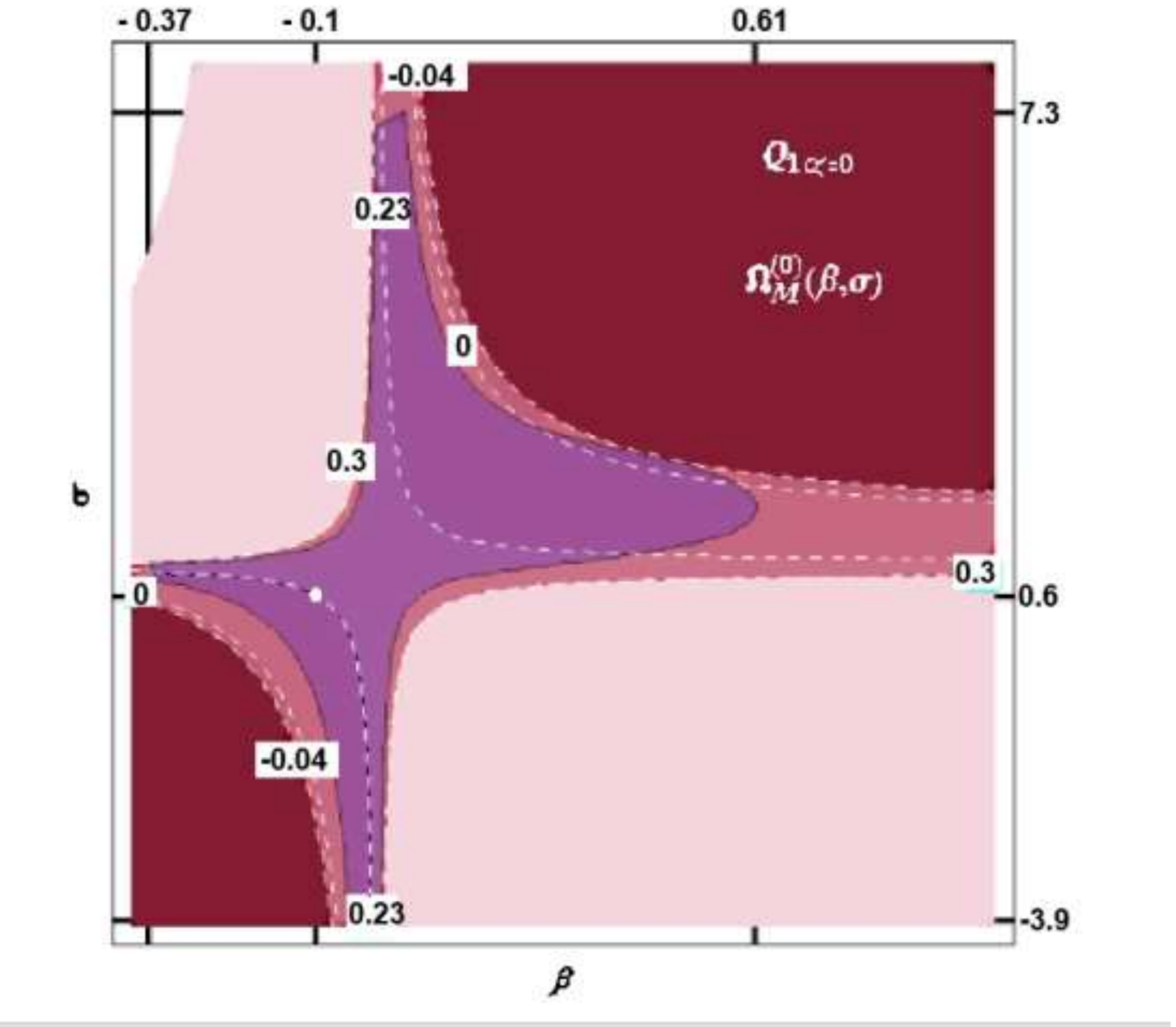}
\caption{\scriptsize{ $1\sigma$ CL for the present matter content in the $\beta$\,  vs. $\sigma$  parameter space for the best fit model with the interaction $\mathcal{Q}_{1_{\alpha=0}}$. The restriction on the possible values of $\Omega_1$ plus $\Omega_m$ is obtained by overlapping the confidence region $1\sigma$ of plane $\beta$~vs. $\sigma$ (lilac region) on contours of the total density parameter of matter $\Omega_{M0}$. At $1\sigma$ CL, the present values of  $\Omega_{M0}$ belong to the interval $[0,0.3]$ with the best fit value 0.23 (white dot). }}
\label{Fig:.Alpha0Fig4}
\end{minipage}
\end{figure}

\begin{figure}
\centering 
\begin{minipage}[t][12cm][b]{0,5\textwidth}
\includegraphics[height=15cm,width=10cm]{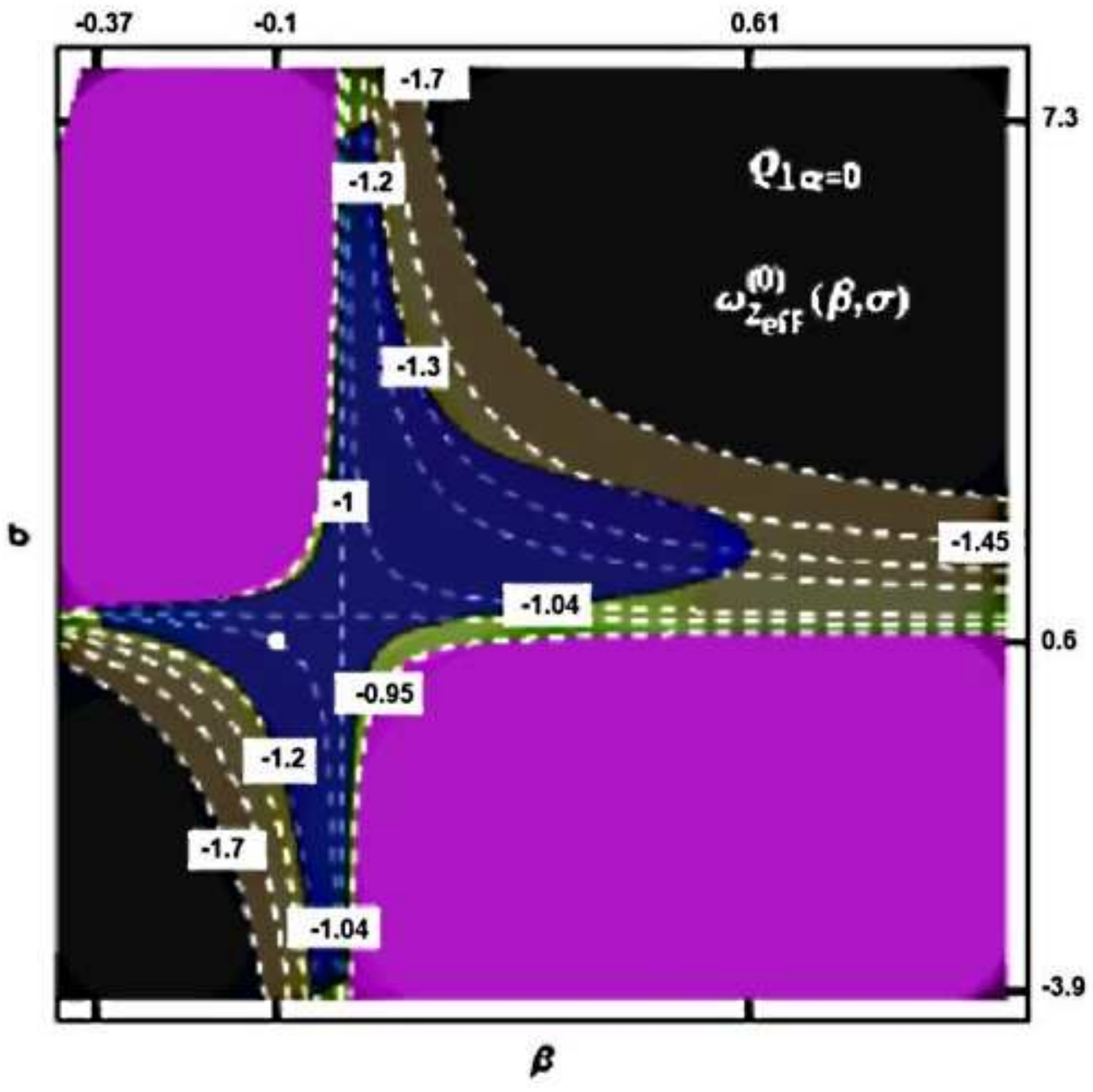}
\caption{\scriptsize{ $1\sigma$ CL for the actual effective dark energy EoS in the $\beta$\,  vs. $\sigma$  parameter space for the best fit model with the interaction $\mathcal{Q}_{1_{\alpha=0}}$. The restriction on the possible values of $\omega_{2eff}$ is obtained by overlapping the confidence region $1\sigma$ of plane $\beta$~vs. $\sigma$ (blue region) on contours of the effective  equation of state of the DE density (dashed curves). There it can be seen that at $1\sigma$ CL, the present values of  $\omega_{2eff}$ belong to the interval $[-0.95,-1.45]$ with the best fit value -1.04 (white dot). }}
\label{Fig:.Alpha0Fig5}
\end{minipage}
\end{figure}

\begin{figure}
\centering 
\begin{minipage}[t][12cm][b]{0,5\textwidth}
\includegraphics[height=15cm,width=10cm]{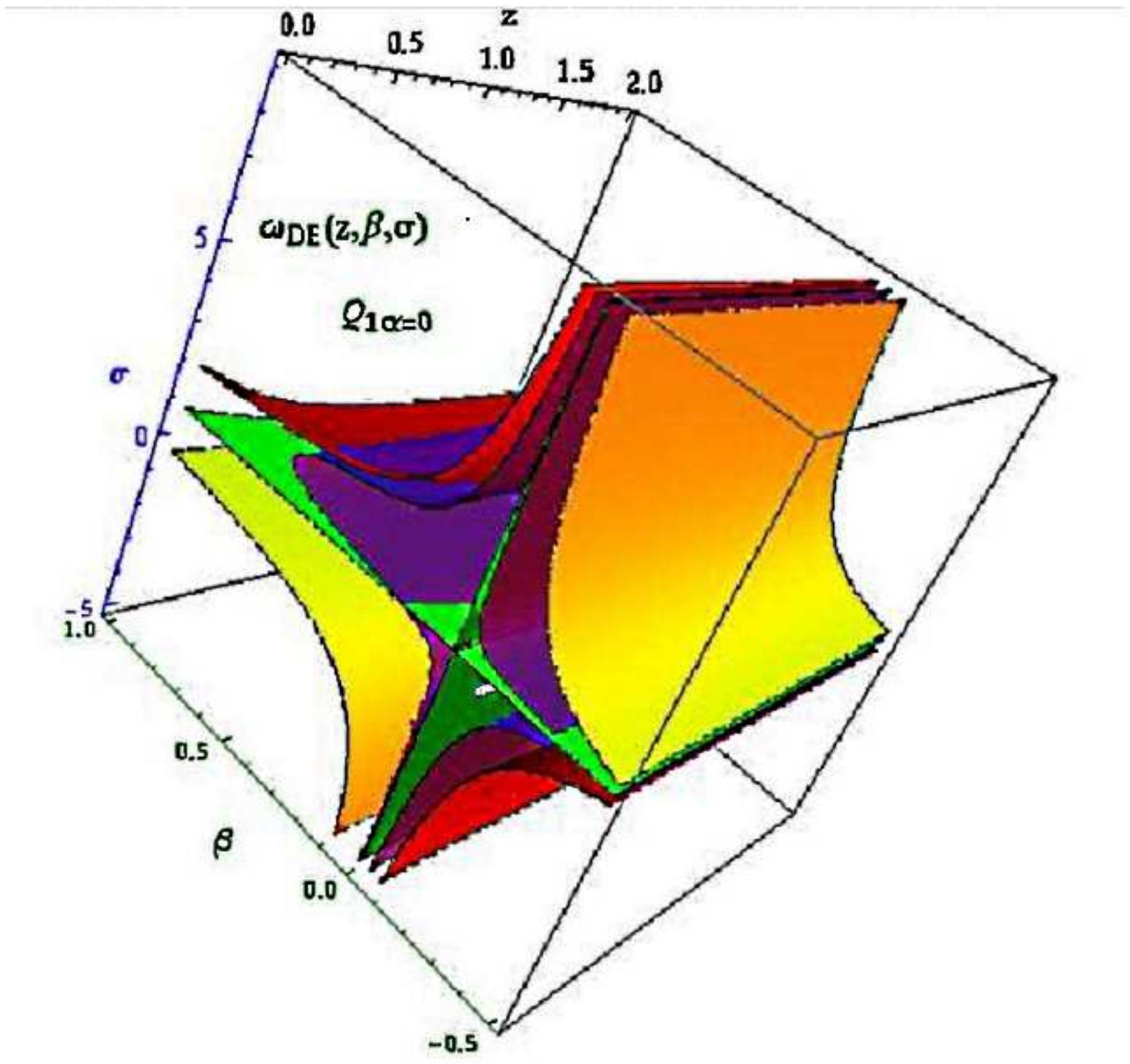}
\vspace{-0.6cm}
\caption{\scriptsize{Evolution of the dark energy effective equation of state $\omega_{2eff}(z,\beta,\sigma)$ at  $1\sigma$ CL in the $\beta$\,  vs. $\sigma$  parameter space for the interaction $\mathcal{Q}_{1_{\alpha=0}}$. The best fit white line crosses the phantom divide plane (green plane) $\omega_{\Lambda}=-1$ just before $ z=0 $. It is shown that between $z=0$ and $z=2$, the dark energy fluid EoS is always between the yellow plane $\omega_{DE}=-0.5$ and the red plane $\omega_{DE}=-1.5$ at $1\sigma$ CL. The crossing of the red plane occurs only for  $\beta > 0$ which is excluded because it makes $\rho_2 < 0$.}}
\label{Fig:.Alpha0Fig6}
\end{minipage}
\end{figure}

A graphical way of knowing the range of variation of the current total matter content, caused by changes in $\beta$ and $\sigma$,  is throughout the overlapping of the $1\sigma$ confidence region for these coupling parameters, on the contours graph of  the present matter content $\Omega_{M0}=\Omega_m(0)+\Omega_1(0)$  in the $\beta$\,  vs. $\sigma$  parameter space for the best fit model. The restriction on the possible present values of $\Omega_1$ plus $\Omega_m$ in the case $\mathcal{Q}_{1_{\alpha=0}}$ can be seen in Fig.\ref{Fig:.Alpha0Fig4} where the $1\sigma$ confidence region (lila region) restricts $\Omega_{M0}$ so that its present values  belong to the interval $[0,0.3]$. Also in this two-dimensional graph, we can see the $1\sigma$ CL, $\beta= -0.1_{-0.27}^{+0.71}$, $\sigma=0.6_{-4.5}^{+6.7}$ and the best fit value for the matter parameter  $\Omega_{M0}=0.23$ (white dot).\\
The characteristic parameter of dark energy, its effective equation of state  $\omega_{2eff}$, deserves some further study. Its variation with the parameters that identify the interaction, $\beta$ and $\sigma$, can be seen in Fig.\ref{Fig:.Alpha0Fig5}, where the region corresponding to the $1\sigma$ CL in the parameter space $\beta$~vs. $\sigma$ (blue area), is superimposed on the contour plot of the current $\omega_{2eff}$. The resulting information is that the actual effective dark energy EoS is $\omega_{2eff}=-1.04_{-0.41}^{+0.09}$. For clarity, Fig.\ref{Fig:.Alpha0Fig4} and Fig.\ref{Fig:.Alpha0Fig5} are presented complete although only negative values of $\beta$ are allowed by the positivity of $\rho_2$.   Moreover, from the three dimensional Fig.\ref{Fig:.Alpha0Fig6}, we can see the evolution of $\omega_{2eff}(z)$ between $z=0$ and $z=2$ at $1\sigma$ CL in the  $\beta$~vs. $\sigma$ region. The plot shows that best fit white line crosses the phantom divide plane (green plane) $\omega_{\Lambda}=-1$ just before the current time and that, in that redshift interval, the EoS of the dark energy fluid  is always between the yellow plane $\omega_{DE}=-0.5$ and the red plane  $\omega_{DE}=-1.5$ at least for values of $\beta$, $\sigma$ belonging to their $1\sigma$ CL. 

\vskip 2.5cm
%Alpha0Fig4

\begin{figure}
\centering 
\begin{minipage}[t][12cm][b]{0,5\textwidth}
\includegraphics[height=15cm,width=10cm]{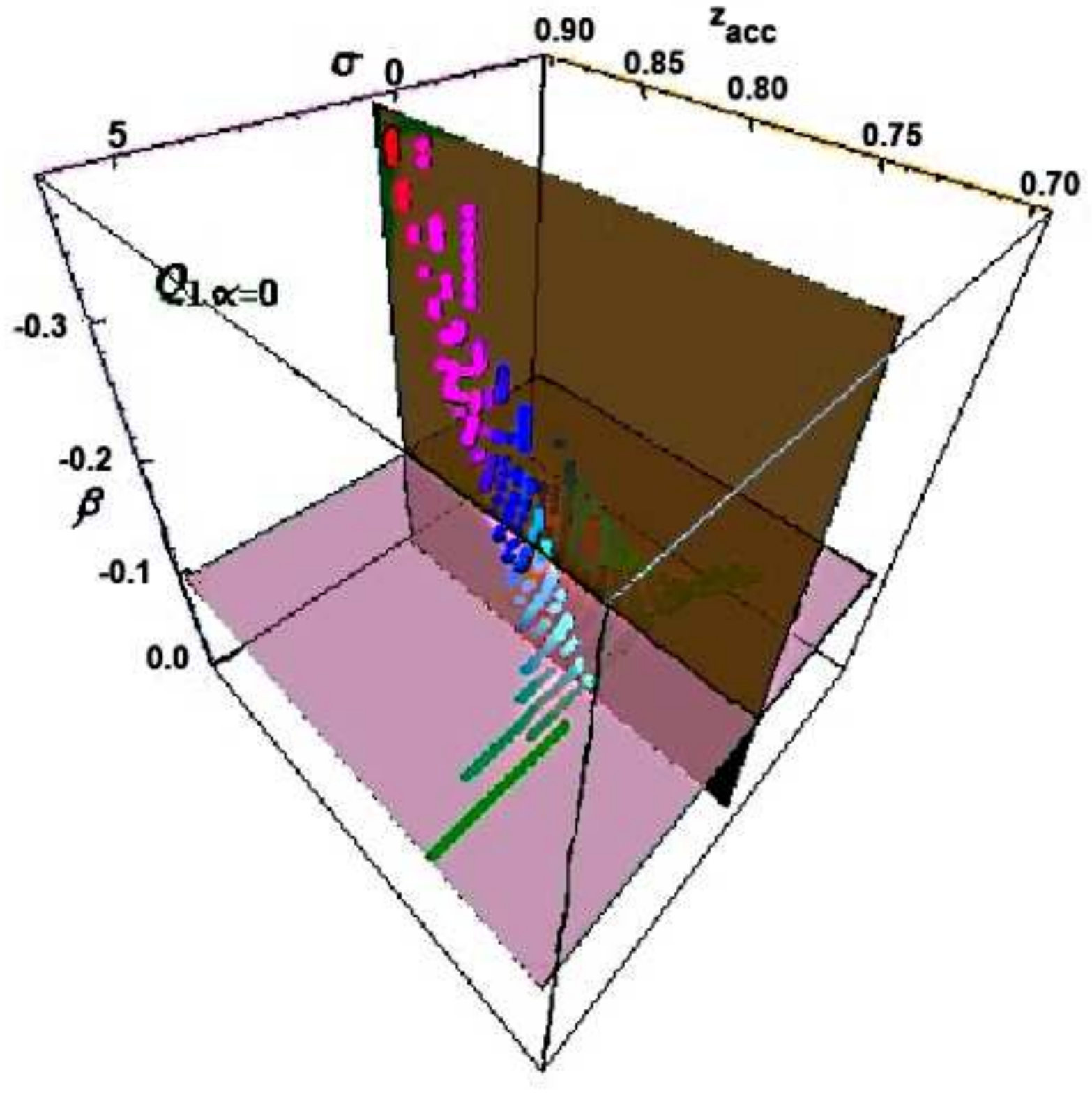}
\caption{\scriptsize{Variation of $z_{acc}$ due to changes at $1\sigma$ CL in the $\beta$\, vs. $\sigma$ parameter space in the case $\mathcal {Q}_{1_{\alpha=0}}$. The transition value $z_{acc}(\beta,\sigma)$ is a decreasing function of $\beta$ and an increasing one of $\sigma$, much more sensitive to changes in $\beta$ than to changes in $\sigma$. The brown constant plane corresponds to the best fit value $\sigma_{bf}=0.6$ and the pink constant plane corresponds to the best fit $\beta_{bf}=-0.1$.}}
\label{Fig:.Alpha0Fig7}
\end{minipage}
\end{figure}

The crossing of the red plane occurs only for $\beta > 0$, but the positive values are excluded for $\beta$ because they lead to $\rho_2 < 0$. This result is in good agreement with the constraints on the evolution of the dark energy equation of state derived from a $WMAP+SNLS+BAO+H(z)$ dataset for several dark energy models as Hu-Sawicki, tracking SUGRA and tracking power-law \cite{Said:2013jxa,Fang:2008sn,Ade:2013zuv}.
The variation of the parameters $\beta$ and $\sigma$ also affects the redshift of the transition to accelerating universe.  Fig.\ref{Fig:.Alpha0Fig7} shows that with variations on the coupling parameters at $1\sigma$ CL, the transition value  $z_{acc} $ belongs to the interval $[0.74,0.89]$ being an increasing function of $\sigma$ and decreasing one of $ \beta$, much more sensitive to changes in $ \beta$ than to changes in $\sigma$ as we can observe on brown constant plane $\sigma_{bf}=0.6$ and on green constant plane  $\beta_{bf}=-0.1$. This best fit value for the transition redshift $z_{acc}=0.805$ is significantly higher than that obtained in \cite{Cunha:2008ja} at $1\sigma$ CL analyzing distinct databases and in a model-independent way.  However, it is typical for models with interactions in the dark sector \cite{Chimento:2007yt,Chimento:2011dw,Chimento:2013se,Chimento:2007da,Forte:2012ww,Farooq:2013hq}.

%Alpha0Fig5

%Alpha0Fig6

%Alpha0Fig7

We can estimate the age of our universe $T$ using the time-redshift relation 
\be
\n{022}
t(z)=\int_z^{\infty}\frac{dx}{(1+x)\sqrt{\rho_t(x)/3}}, \qquad  T=t(0),
\ee
\no and also qualify the interaction, at least with respect to the coincidence problem, through the fraction of T  for which the ratio between dark sector densities remains around the  unity. That is, if for the interaction $\mathcal{Q}_i$, it turns out that from $z^{(\mathcal{Q}_{i})}$ until today is $r(z^{(\mathcal{Q}_{i})})\sim 1$, then the magnitude 
\be
\n{023}
\it{quality}(\mathcal{Q}_{i})=1 - \frac{ t(z^{(\mathcal{Q}_{i})})}{T},
\ee 
\no gives us a good idea of the benefits of the interaction under study respect to that issue. 
In the particular case $\mathcal{Q}_{1_{\alpha=0}}$ with its best fit model, the coincidence seems to be satisfied for $z^{(\mathcal{Q}_{1_{\alpha=0}})} \sim 0.443$ (see Fig.\ref{Fig:.Alpha0Fig2}).  Therefore, replacing (\ref{011}) in (\ref{022}) it turns out that the best fit model has a good score $ \textit{quality}(\mathcal{Q}_{1_{\alpha=0}})=1- t(0.443)/t(0)=33.13\%$ with $T=13.86$  Gyr.

%%%%%%%%%%%%%%
%%%%%%%%%%%%%%%%%

\subsubsection{\bf{$\alpha\ne 0$}}

%%%%%%%%%%%%%%
%%%%%%%%%%%%%%

When the term in $\rho_2'$ is included in the interaction (\ref{06}), the evolution equation for the auxiliary dark barotropic index $\gamma$ is
\be
\begin{aligned}
\n{024}
\textstyle{\gamma'((1+\alpha}&\textstyle{-\alpha\sigma)- \frac{3}{2}\alpha\gamma)}\\
&\textstyle{-\frac{3}{2}\alpha(\gamma-\gamma_1)(\gamma - \gamma_+)(\gamma_- -\gamma)}=\textstyle{0},
\end{aligned}
\ee
\no where
\be
\n{025}
\textstyle{\gamma_{\pm}}=\textstyle{\frac{2(1+\alpha-\alpha\sigma)+3\beta \mp \sqrt{(2(1+\alpha-\alpha\sigma)+3\beta)^2-24\alpha(\gamma_2+\beta-\beta\sigma)}}{6\alpha}}.
\ee
\vskip.2cm

The general solution of (\ref{024}) can be written as
\be
\n{026}
\left(\frac{\gamma_1-\gamma}{\gamma-\gamma_-}\right)^A \left(\frac{\gamma_+-\gamma}{ \gamma-\gamma_-}\right)^B = \bar B_0 e^{3\alpha\eta},
\ee
where $\bar B_0$ is a constant of integration,\\
\vskip.15cm 
$A=\frac{2(1+\alpha-\alpha\sigma)-3\alpha \gamma_1}{(\gamma_- -\gamma_1)(\gamma_1 - \gamma_+)}$ and  $B=\frac{2(1+\alpha-\alpha\sigma)-3\alpha \gamma_+}{(\gamma_1- \gamma_+)(\gamma_+ - \gamma_-)}$.   
\vskip.5cm

Looking at the solution (\ref{026}) for $\gamma(\eta)$, it is apparent that an explicit expression for $\gamma$, and subsequently for $\rho \sim e^{-\int \gamma\ d\eta}$  can be obtained only in special situations, where A or B are canceled, or are inverses one of each other. All these special cases correspond to choosing a particular value for the coupling constant $\alpha$ of the interaction, $ 2(1+\alpha-\alpha\sigma)=3\alpha\gamma_i $ for $ i = 1,+,- $ respectively. 

%%%%%%%%%%%%%%%%%%%%%%%%%%%%%%%%%%%%%%%
%%%%%%%%%%%%%%%%%%%%%%%%%%%%%%%%%%%%%%%%%%%
%%%%%%%%%%%%%%%%%%%%%%%%%%%%%%%%%%%%%%%%%
\begin{itemize}
\item{\bf{$A = 0$}}
\\
 In this case $\alpha$ is fixed at the value $ \alpha=2/(3\gamma_1 + 2(\sigma - 1))$  and therefore the explicit  expressions for $\gamma$ and $\rho$ are
\ben
\n{027}
\gamma(\eta)=\frac{(1-c_2)\gamma_+ + c_2\gamma_-e^{-\eta(\gamma_--\gamma_+)}}{(1-c_2)+ c_2e^{-\eta(\gamma_--\gamma_+)}},
\een
\ben
\n{028}
\rho(z)=\bar \rho_0\left(c_2(1+z)^{3\gamma_-}+(1-c_2)(1+z)^{3\gamma_+}\right),
\een
\no where $0\leq c_2 \leq 1$ and $\bar\rho_0$ are constants of integration. 
From equations (\ref{03}), (\ref{027}) and (\ref{028}) we obtain the explicit expression of all interesting variables of the case, the total energy density $\rho_t$, the partial energy densities $\rho_1$ and $\rho_2$, and the dark ratio $r=\rho_1/\rho_2$,

\ben
\n{029}
\textstyle{\frac{\rho^{(\mathcal{Q}_{1_{A=0}})}_t(z)}{\rho_0}}=\textstyle{\Big[b_mx^{\gamma_m}+(1-b_m)\Big(c_2 x^{\gamma_-}+(1-c_2) x^{\gamma_+}\Big)\Big]},\ \ \ 
\een

\ben
\n{030}
\textstyle{\frac{\rho^{(\mathcal{Q}_{1_{A=0}})}_1(z)}{\bar\rho_0}}= \textstyle{\frac{\Big(c_2 (\gamma_- -\gamma_2)x^{\gamma_-}+(1-c_2)(\gamma_+-\gamma_2) x^{\gamma_+}\Big)}{(\gamma_1-\gamma_2)}}, 
\een

\ben
\n{031}
\textstyle{\frac{\rho^{(\mathcal{Q}_{1_{A=0}})}_2(z)}{\bar\rho_0}}= \textstyle{\frac{\Big(c_2 (\gamma_1 -\gamma_-)x^{\gamma_-}+(1-c_2)(\gamma_1-\gamma_+) x^{\gamma_+}\Big)}{(\gamma_1-\gamma_2)}}, 
\een
\no and

\ben
\n{032}
\textstyle{r^{(\mathcal{Q}_{1_{A=0}})}(z)}=\textstyle{\frac{c_2(\gamma_- - \gamma_2)x^{\gamma_-} + (1- c_2)(\gamma_+ - \gamma_2)x^{\gamma_+}}{c_2(\gamma_1 - \gamma_-)x^{\gamma_-} + (1- c_2)(\gamma_1 - \gamma_+)x^{\gamma_+}}},
\een

\no where $x=(1+z)^3$, $b_m=\frac{\rho_{0m}}{\rho_0}$, $\frac{\bar\rho_0}{\rho_0}=1-b_m$  and $\rho_0$ is the actual total energy density.

Also, we obtain the explicit expressions for the interaction $\mathcal{Q}_{1_{A=0}}$, the corresponding deceleration parameter $q$, the density parameter for the dark energy $\Omega_2$, and the effective EoS $\omega_{1eff}$, $\omega_{2eff}$ and $\omega_{t}$,

\begin{multline}
\n{033}
\textstyle{\frac{\mathcal{Q}_{1_{A=0}}(z)}{\rho_0(1-b_m)}}= \textstyle{\frac{-\Big(c_2\nu_- x^{\gamma_-}+(1-c_2)\nu_+ x^{\gamma_+}\Big)}{(\gamma_1-\gamma_2)\nu_1\Big(c_2 x^{\gamma_-}+(1-c_2)x^{\gamma_+}\Big)}}   \times \ \ \ \ \  \ \ \ \ \ \\ 
 \textstyle{\big(c_2(\gamma_1-\gamma_-)(\gamma_- - \frac{\beta\nu_1}{2}) x^{\gamma_-}+(1-c_2)(\gamma_1-\gamma_+)(\gamma_+ - \frac{\beta\nu_1}{2}) x^{\gamma_+}\big)},\\ 
\end{multline}

\be
\n{034}
\textstyle{1+q^{(\mathcal{Q}_{1_{A=0}})}(z)}=\textstyle{\frac{3}{2} \frac{b_m\gamma_mx^{\gamma_m}+(1-b_m)(c_2 \gamma_- x^{\gamma_-}+(1-c_2)\gamma_+ x^{\gamma_+})}{b_mx^{\gamma_m}+(1-b_m)\Big(c_2 x^{\gamma_-}+(1-c_2) x^{\gamma_+}\Big)},}\ \ 
\ee

\be
\n{035}
\textstyle{\Omega_2^{(\mathcal{Q}_{1_{A=0}})}(z)}=\textstyle{\frac{(1-b_m)}{\gamma_1-\gamma_2}   \frac{c_2 (\gamma_1-\gamma_-) x^{\gamma_-}+(1-c_2)(\gamma_1-\gamma_+) x^{\gamma_+}}{b_mx^{\gamma_m}+(1-b_m)\Big(c_2 x^{\gamma_-}+(1-c_2) x^{\gamma_+}\Big)},}
\ee

\be
\begin{aligned}
\n{036}
&\omega_{1eff}^{(\mathcal{Q}_{1_{A=0}})}(z)=\textstyle{\omega_{1} - \frac{ \Big(c_2\nu_-x^{\gamma_-}+(1-c_2)\nu_+x^{\gamma_+}\Big)}{\nu_1\Big(c_2 x^{\gamma_-}+(1-c_2) x^{\gamma_+}\Big)}} \times \\
&\textstyle{ \frac{\Big(c_2(\gamma_1-\gamma_-)(\gamma_- - \frac{\beta}{2}\nu_1)x^{\gamma_-}+ (1-c_2)(\gamma_1-\gamma_+)(\gamma_+ - \frac{\beta}{2}\nu_1) x^{\gamma_+}\Big)}{\Big(c_2(\gamma_--\gamma_2)x^{\gamma_-}+(1-c_2)(\gamma_+-\gamma_2) x^{\gamma_+}\Big)},}
\end{aligned}
\ee

\be
\begin{aligned}
\n{037}
&\omega_{2eff}^{(\mathcal{Q}_{1_{A=0}})}(z)=\textstyle{\omega_{2} + \frac{ \Big(c_2\nu_- x^{\gamma_-}+(1-c_2) \nu_+x^{\gamma_+}\Big)}{\nu_1\Big(c_2 x^{\gamma_-}+(1-c_2) x^{\gamma_+}\Big)}} \times \\
&\textstyle{ \frac{\Big(c_2(\gamma_1-\gamma_-)(\gamma_- - \frac{\beta}{2}\nu_1)x^{\gamma_-}+ (1-c_2)(\gamma_1-\gamma_+)(\gamma_+ - \frac{\beta}{2}\nu_1) x^{\gamma_+}\Big)}{\Big(c_2(\gamma_1-\gamma_-)x^{\gamma_-}+(1-c_2)(\gamma_1-\gamma_+) x^{\gamma_+}\Big)},}
\end{aligned}
\ee

\be
\begin{aligned}
\n{038}
\textstyle{\omega_t^{(\mathcal{Q}_{1_{A=0}})}(z)=\frac{\Big[\omega_mb_mx^{\gamma_m}+(1-b_m)\Big(\omega_-c_2 x^{\gamma_-}+\omega_+(1-c_2) x^{\gamma_+}\Big)\Big]}{\Big[b_mx^{\gamma_m}+(1-b_m)\Big(c_2 x^{\gamma_-}+(1-c_2) x^{\gamma_+}\Big)\Big]}}
\end{aligned}
\ee

\no where $\nu_i=3\gamma_i+2(\sigma-1)\ \ \ i=1,-,+ $ and $\omega_j=\gamma_j-1$ with $j=m,-,+$.

Here, the best-fit model through the minimization of the function $\chi^2$ (\ref{020}), where the theoretical Hubble function $H_{th}=\sqrt{\rho_t/3}$ is taken from the equation (\ref{016}), corresponds to a minimun $\chi^2_{min}=17.7573$ or $\chi^2_{dof}=0.845586$. The set of parameters is $\Theta_{\mathcal{Q}_{1_{A=0}}}=(H_0, b_m, c_2, \gamma_m, \gamma_1, \gamma_2, \beta, \sigma)$ with the best-fit values $H_0=71.42\rm{kms^{-1}Mpc^{-1}}$, $b_m=0.04$, $c_2=0.45$, $\gamma_m=1$, $\gamma_1=1.2$, $\gamma_2=0$, $\beta=-0.745$ and $\sigma=0.22$. With these values we can say that the model belongs to a scenario of warm dark matter (WDM) fluid interacting with a cosmological constant in presence of  non-interactive dust. Conveniently, the best fit value for $\beta$ is negative, which keeps positive the energy density of DE at any time in the past. The nickname WDM is used here in the following sense. The perfect gas equation of state may be written as $p = \rho_tRT$ with the particular gas constant  R and the temperature T. $C=\sqrt{RT}$ is a characteristic thermal speed of the molecules and if $c$ is the speed of light, the EoS  $\omega_t = \frac{p_t}{c^2\rho_t} =\frac{C^2}{c^2}$  is a measure of the energy of the matter directly related to its typical temperature \cite{Carter:1987qr}. In this sense, we call lukewarm or generically, warm matter, (baryonic or dark), to that has a lightly positive EoS ($0<\omega_t<1/3$), reserving the expression  ``cold matter" for the case of EoS strictly zero.  
The issue of the dark matter component with a small but non zero pressure was studied in \cite{Harko-Lobo:2012,Bharadwaj-Kar:2003,Su-Chen:2009,Saxton-Ferreras:2010,Lim:2010yk} in the context of modeling galaxy halos. Also in \cite{Moore:1999gc,Bode:2000gq,AvilaReese:2000hg,Dalcanton:2000hn,Boyanovsky:2010pw,Destri:2012yn},
WDM particles were invoked as possible solutions to both, the over prediction of satellite galaxies, by almost an order of magnitude larger than the number of satellites that have been observed in Milky-Way sized galaxies and as a mechanism to smooth out the cusped density profiles predicted by CDM simulations.

%A0Fig1
\begin{figure}
\vskip -2.5cm
\centering 
\begin{minipage}[t][12cm][b]{0,48\textwidth}
\includegraphics[height=22cm,width=18cm]{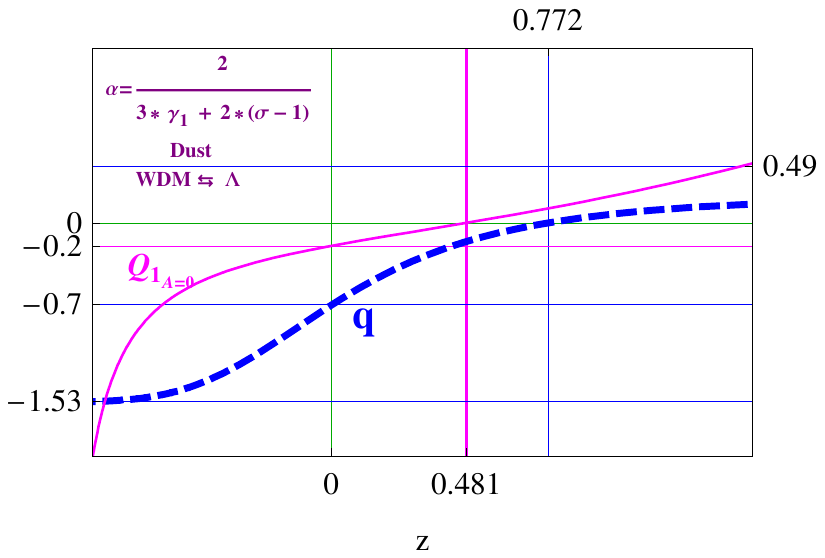}
\caption{\scriptsize{Evolutions of the interaction (\ref{033}) and its corresponding deceleration parameter $q$ (\ref{034}), in the case $\small{\mathcal{Q}_{1_{A=0}} =(3\gamma+2(\sigma-1))(\rho_2'+\beta\nu_1\rho_2/2)/\nu_1}$ for the best fit model ($\chi^2_{dof}=0.845586$), with $H_0=71.42~{\rm km~s^{-1}\,Mpc^{-1}}$, $b_m=0.04$, $\gamma_m=1$, $\gamma_1=1.2$, $\gamma_2=0$, $\beta=-0.745$,  $\sigma=0.22$,   and $c_2= 0.45$. The transition to the accelerated regime is verified at $z_{acc}=0.772$ and the actual deceleration parameter value is $q_0=-0.7$. The interaction changes its sign at $z_{\mathcal{Q}_{1_{A=0}}}=0.481$, from which begins to transfer energy from DE to DM. The gradient of this interaction between the present epoch and the time when it reverses its sign, is an order of magnitude greater than in the case of $\alpha = 0$.}}
\label{Fig:.A0Fig1}
\end{minipage}
\end{figure}

%A0Fig2
\begin{figure}
\vskip -2.8cm
\centering 
\begin{minipage}[t][12cm][b]{0,48\textwidth}
\includegraphics[height=24cm,width=15cm]{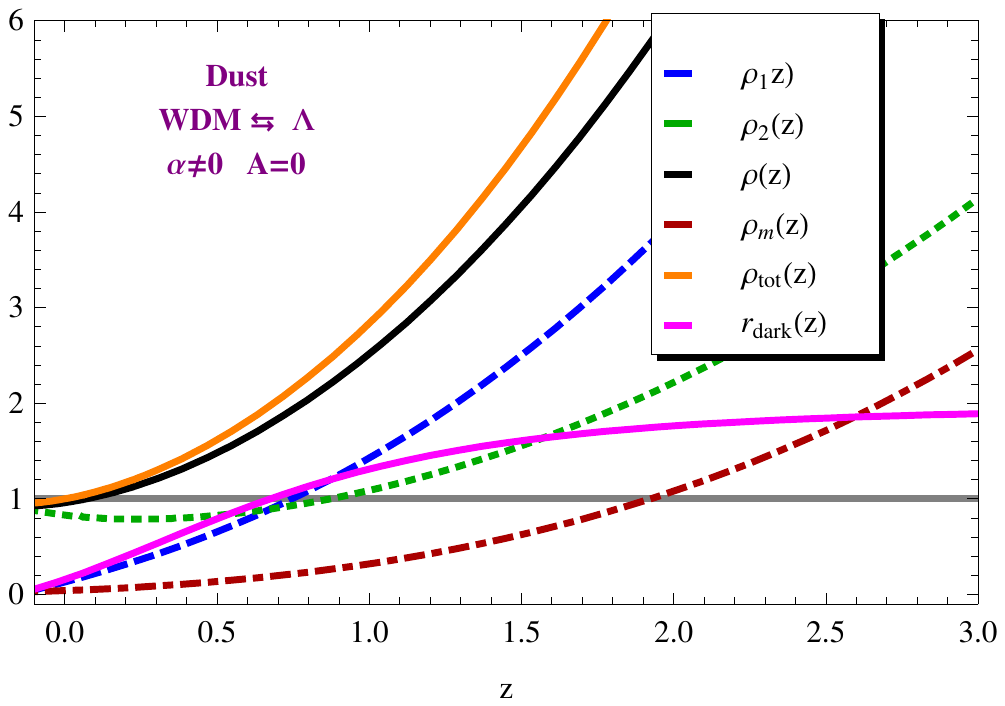}
\vspace{-0.6cm}
\caption{\scriptsize{Evolution of all energy densities involved in the best fit model with interaction $\mathcal{Q}_{1_{A=0}}$ in units of $3H_0^2$. Interestingly, most of the total energy density is supplied by the dark sector throughout the evolution. The ratio of dark energy densities becomes  of order 1 around $z = 0.7$, much earlier than in the previous case $\alpha = 0$ and  this is extended up well back in time.}}
\label{Fig:.A0Fig2}
\end{minipage}
\end{figure}

In Fig.:\ref{Fig:.A0Fig1}, the interaction $\mathcal{Q}_{1_{A=0}}$ and the corresponding deceleration parameter $q^{(\mathcal{Q}_{1_{A=0}})}$ are depicted as functions of the redshift. In this model there is strong energy transfer from the particles of warm dark matter to DE at early times and the situation is equally serious, but in the opposite direction, in the distant future.  Instead, in recent history the transfer is relatively smooth, only one order of magnitude greater than for $\alpha = 0$, and WDM particles decay into DE until  $z_{\mathcal{Q}}=0.481$ when its effects are reversed causing the creation of WDM particles. For $z >>1$, $q$ tends asymptotically to a  value $q_e=0.49$ very similar to the $\Lambda$CDM at early times. The same is true for the current time when  $q_0=-0.7$ but in the far future this model is a fifty percent more accelerated. The Fig.:\ref{Fig:.A0Fig2} shows the evolution of all  energy densities $\rho_1$, $\rho_2$, $\rho_m$, $\rho$ and $\rho_t$ for the best fit model and also the ratio $r=\rho_1/\rho_2$ for this particular interaction. There it can be seen that the full coincidence between the magnitudes of the dark densities  $r=1$ occurs around $z\sim 0.7$, before than in the previous case $\alpha=0$, and remains in the same order until well back in time.  A notable feature is that the total energy is mainly composed of energy of the interacting dark sector, and that it varies much more violently than in the previous case. The evolutions of all effective EoS, whose early - time asymptotic values are $\omega_{1eff}=0.44$, $\omega_{2eff}= -1.47$  and $\omega_{t}=0$, are shown in Fig.:\ref{Fig:.A0Fig3}. The effective EoS of DE crosses the PDL around $z=0.48$ but, notably, coming from more negative values. This crossing is treated in a lot of works \cite{QuintomsFeng:2004ad} and particularly, it is considered in \cite{HutererCooray} where a passage from the value $-1.3$ at $z = 1.25$ to the value $-0.6$ at $z = 0.3$  is obtained through the Gold set of SNe Ia. The actual EoS are $\omega_{t}(0)=-0.8$, $\omega_{1eff}(0)=-1.315$ and $\omega_{2eff}(0)=-0.763$.

%A0Fig3
\begin{figure}
\vskip -3.2cm
\centering 
\begin{minipage}[t][13cm][b]{0,48\textwidth}
\includegraphics[height=32cm,width=20cm]{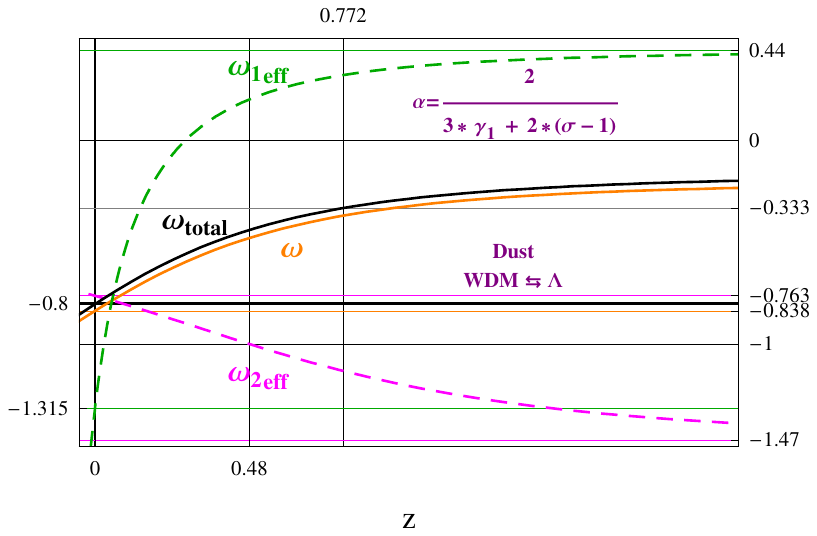}
\vspace{-0.6cm}
\caption{\scriptsize{Evolution of all EoS in the best fit model with the interaction $\mathcal{Q}_{1_{A=0}}$. At early time the global effective fluid behaves as dust while the asymptotic effective values of the dark EoS are $\omega_{1eff}=0.44$ and $\omega_{2eff}=-1.47$. The crossing of PDL is carried out by the effective DE around $z=0.48$, while the actual values of the effective EoS are $\omega_{total}=-0.8$, $\omega_{1eff}=-1.315$ and $\omega_{2eff}=-0.763$.}}
\label{Fig:.A0Fig3}
\end{minipage}
\end{figure}

The current matter content (WDM plus baryonic cold matter) is graphically constrained in  Fig.:\ref{Fig:.A0Fig4} where confidence regions at level $1\sigma$ and $2\sigma$ for coupling coefficients $\beta$ and $\sigma$ are plotted on the contours of $\Omega_{M0}=(\rho_1(0)+\rho_m(0))/\rho_t(0)$. The permissible values for total matter content correspond to the range $[0,0.24]$ at $1\sigma$ CL in $\beta$ vs. $\sigma$ space and $[0,0.3]$ at $2\sigma$ CL while the best fit is $\Omega_{M0}=0.17$. These ranges are in good agreement with the results found in the literature \cite{Kessler:2009ys,Sullivan:2011kv,Suzuki:2011hu}.

%A0Fig4
\begin{figure}
\vskip -3.2cm
\centering 
\begin{minipage}[t][13cm][b]{0,48\textwidth}
\includegraphics[height=25cm,width=19cm]{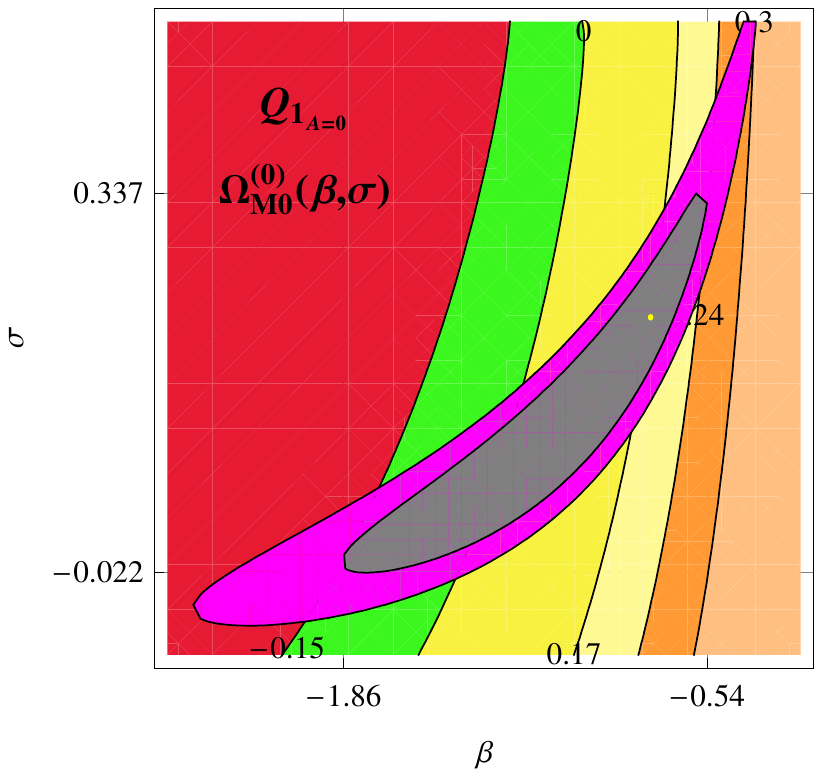}
\vskip-0.5cm
\caption{\scriptsize{$1\sigma$ and $2\sigma$ confidence regions for the parameter space $\beta$ vs. $\sigma$ overlapping  the sum $\Omega_{M0}$ of the two actual density parameters of matter (WDM and baryonic cold matter) in the case $\mathcal{Q}_1$ with $A=0$. From the contours of the density parameter we can infer that non-negative values, that is, the permissible values for total matter content correspond to the range $[0,0.24]$ at $1\sigma$ CL and $[0,0.3]$ at $2\sigma$ CL while the best fit is $\Omega_{M0}=0.17$.}}
\label{Fig:.A0Fig4}
\end{minipage}
\end{figure}
With the same graphical technique, the effective EoS of DE at the present time is constrained giving the range $[-0.91,0.735]$ at $1\sigma$ CL in $\beta$ vs. $\sigma$ space, as seen in  Fig.:\ref{Fig:.A0Fig5}. From these contours  one can infer that the value of $\omega_{2eff}(0)=-1$ is not favored even at $2\sigma$ CL in $\beta$ vs. $\sigma$ space. 

%A0Fig5
\begin{figure}
\vskip -3.2cm
\centering 
\begin{minipage}[t][13cm][b]{0,48\textwidth}
\includegraphics[height=25cm,width=19cm]{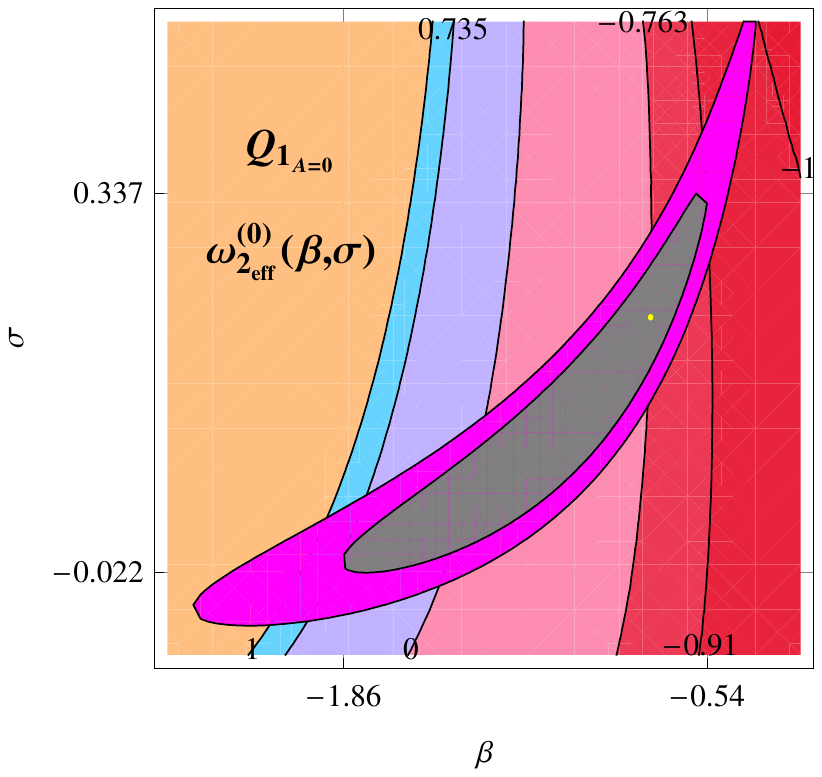}
\vskip-0.5cm
\caption{\scriptsize{$1\sigma$ and $2\sigma$ confidence regions for the parameter space $\beta$ vs. $\sigma$ overlapping  the actual effective EoS of dark energy in the case $\mathcal{Q}_{1_{A=0}}$. From the contours of $\omega_{2eff}(0)$ one can infer that for variations of $\beta$ and $\sigma$ at $1\sigma$ CL, the values of the actual effective dark energy EoS belong to the range $[-0.91,0.735]$. The  value $\omega_{2eff}(0)=-1$ is not favored even at $2\sigma$ CL,  while the best fit is $\omega_{2eff}=-0.763$.}}
\label{Fig:.A0Fig5}
\end{minipage}
\end{figure}

With the three-dimensional graph of  Fig.:\ref{Fig:.A0Fig6}, obtained by intersecting the $1\sigma$ CL ellipses  in each instant between $z=0$ and $z=2$ (yellow surface), on level surfaces $\omega_{2eff}(z)$, we can extend  the above analysis to  study the evolution of these constraints in the redshift range $[0,2]$. The  contours considered are $ \omega_{2eff}=-1.5$ (red surface), $ \omega_{2eff}=-1$ (green surface) and $ \omega_{2eff}=-0.5$ (light blue surface). The pink line, which represents the evolution of $ \omega_{2eff}$ for the best fit  model, does not cross the light blue surface when $z=0$. Instead at $1\sigma$ CL in $\beta$ vs. $\sigma$ space, $\omega_{2eff}(z=0)$ yes it does  and at $1\sigma$ CL the crossing of the PDL is carried out between $z=0.3$ and $z=0.5$. Again,  this is in agreement with literature \cite{Nesseris:2006er}.\\
%A0Fig6
\begin{figure}
\centering 
\begin{minipage}[t][13cm][b]{0,48\textwidth}
\includegraphics[height=18cm,width=10cm]{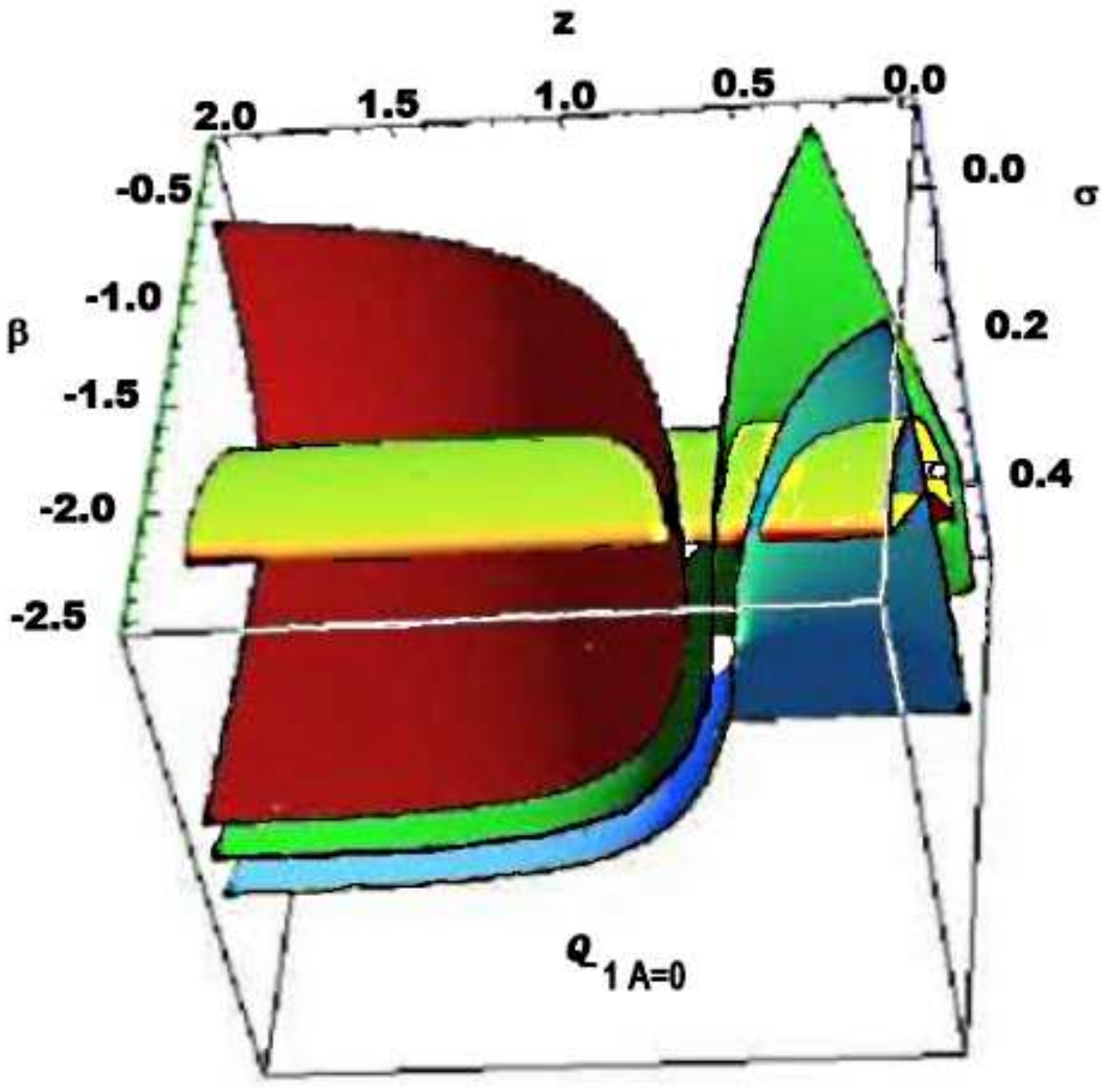}
\caption{\scriptsize{Evolution at $1\sigma$ CL of the effective EoS of dark energy in the case $\mathcal{Q}_{1_{A=0}}$. The figure is obtained by superposition of the $1\sigma$ CL surface (yellow surface) for the space $\beta$ vs. $\sigma$ vs. $z$ and the  contours $ \omega_{2eff}=-1.5$ (red surface), $ \omega_{2eff}=-1$ (green surface) and $ \omega_{2eff}=-0.5$ (light blue surface). It can be seen that the best fit white line does not cross the light blue surface when $z=0$ (but at $1\sigma$ CL $\omega_{2eff}(z=0)$ yes it does), and that at $1\sigma$ CL the crossing of the PDL is carried out between $z=0.3$ and $z=0.5$.}}
\label{Fig:.A0Fig6}
\end{minipage}
\end{figure}

%A0Fig7
\begin{figure}
\vskip -3.2cm
\centering 
\begin{minipage}[t][13cm][b]{0,48\textwidth}
\includegraphics[height=22cm,width=16cm]{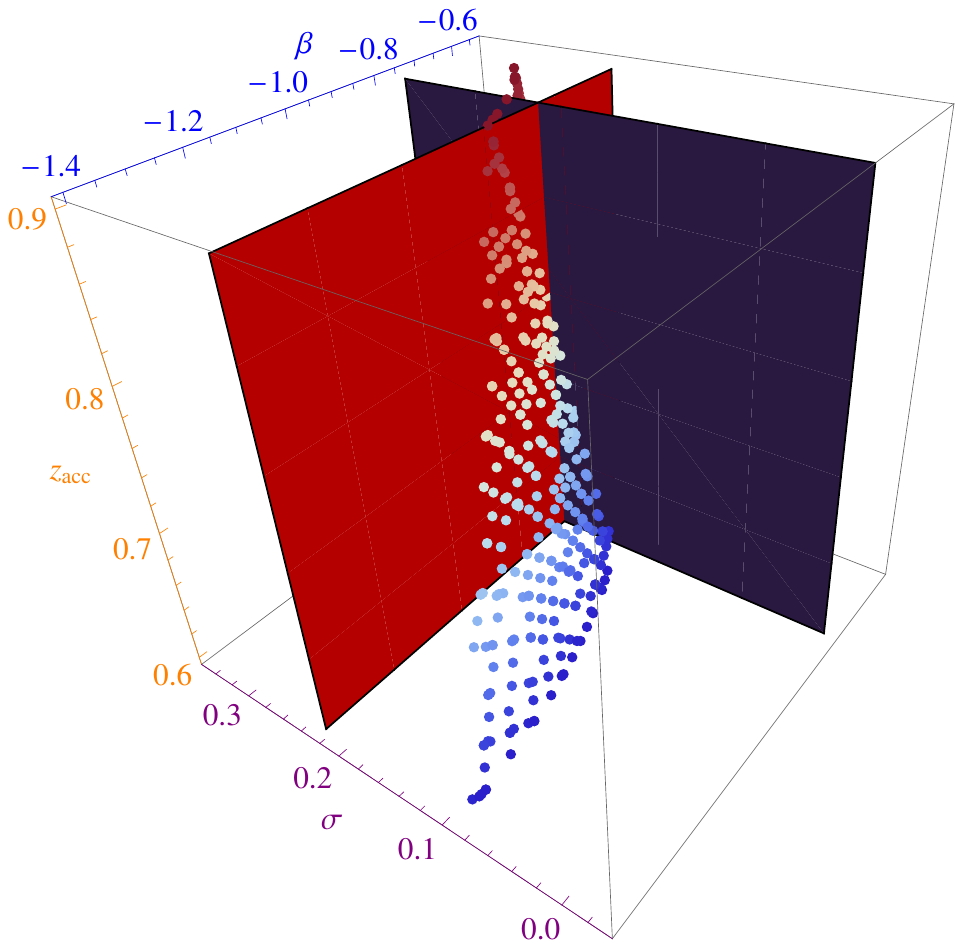}
\caption{\scriptsize{Variation of the transition redshift $z_{acc}$ in the best fit model with the interaction $\mathcal{Q}_{1_{A=0}}$, at $1\sigma$ CL on parameter space $\beta$ vs. $\sigma$. The constant black plane corresponds to $\beta=-0.745$ and the constant red plane to $\sigma=0.22$ while the dots hold the conditions $ \omega_{total}=-1/3$, $\beta\  \epsilon\   [-1.86,-0.54]$ and $\sigma\  \epsilon\   [-0.022,0.337]$. The relation $z_{acc}(\beta,\sigma)$ is an increasing function of $\sigma$  and a decreasing one of $\beta$. }}
\label{Fig:.A0Fig7}
\end{minipage}
\end{figure}

The three-dimensional Fig.:\ref{Fig:.A0Fig7} represents the variation of the transition redshift $z_{acc}$ in the case with interaction $\mathcal{Q}_{1_{A=0}}$, due to variations at $1\sigma$ CL on parameters $\beta$ and $\sigma$. The constant black plane corresponds to $\beta=-0.745$ and the constant red plane to $\sigma=0.22$ while the dots hold the conditions $ \omega_{total}=-1/3$, $\beta\  \epsilon\   [-1.86,-0.54]$ and $\sigma\  \epsilon\   [-0.022,0.337]$. In this context, the relation $z_{acc}(\beta,\sigma)$ is an increasing function of $\sigma$  and a decreasing one of $\beta$.
The best fit model for the case $\mathcal{Q}_{1_{A=0}}$ has $z^{(\mathcal{Q}_{1_{A=0}})} \sim 0.7$.  Therefore, replacing (\ref{029}) in (\ref{022}) it turns out that this interaction has $ \textit{quality}(\mathcal{Q}_{1_{A=0}})=1- t(0.7)/t(0)=43.63\%$, better than the previous coupling $\mathcal{Q}_{1_{\alpha=0}}$ and its cosmological age is $T=14.47$  Gyr.

%%%%%%%%%%%%%%%%%%%%%%%%%%%%%%%%%%%%%%%
%%%%%%%%%%%%%%%%%%%%%%%%%%%%%%%%%%%%%%%%%%%%%

\vskip0.5cm
\item{\bf{$B=0$ or $B=-A$}}
\vskip0.5cm

When any of the options $2(1+\alpha-\alpha\sigma)=3\alpha \gamma_{\pm}$  is satisfied,  the condition $\beta =  \alpha\gamma_2 $ arises from the equation (\ref{025}) and then, equations (\ref{04b}) and (\ref{06}) say us that this condition implies that $\gamma$ is a constant, $\gamma \equiv \gamma_c = 2(1+\alpha-\alpha\sigma)/(3\alpha)$. That is, $\gamma_c = \gamma_{\pm}=\gamma_2$. 
The interactions that satisfy the requirement $\beta =  \alpha\gamma_2$ have $\rho_1=0$, and then they are not of our interest because they do not solve the problem of the coincidence.

\end{itemize}

%%%%%%%%%%%%%%%%%%%%%%%%%%%%%%%%%%%%%%%

\subsection{$3H\mathcal{Q}_2 = 3H\frac{\sigma}{1+\alpha}(\rho_2-\alpha\rho_1)$}

This interaction is the natural extension of the Sun$\&$ Yue proposal, which is recovered when $\alpha=1$ and $\sigma=(1+\alpha)\sigma_{Sun}$, and leads to the first order  equation

\ben
\n{53}
\textstyle{\gamma' - \gamma^2 + (\gamma_1+\gamma_2 - \sigma)\gamma - \gamma_1\gamma_2 +  \frac{\gamma_1+\alpha\gamma_2}{1+\alpha}\sigma=0}.
\een

The general solution of (\ref{53}) 

\ben
\n{54}
\textstyle{\gamma(\eta) = \frac{ c_3\Gamma^-e^{-\eta\Gamma^-}+(1-c_3)\Gamma^+e^{-\eta\Gamma^+}}{c_3e^{-\eta\Gamma^-}+(1-c_3)e^{-\eta\Gamma^+}}}
\een
\no with the constants
\ben
\n{55}
\textstyle{\Gamma^{(\pm)} = \frac{\gamma_1+\gamma_2 - \sigma \pm \sqrt{(\gamma_1+\gamma_2 - \sigma)^2 - 4 (\gamma_1\gamma_2 - \frac{\gamma_1+\alpha\gamma_2}{1+\alpha}\sigma)}}{2},}
\een
allows to obtain the general solution of $d\ln(\rho)/d\eta=-\gamma$ for the energy density of the dark sector $\rho$,
\ben
\n{56}
\textstyle{\rho(z) = \bar\rho\Big(c_3 (1+z)^{3\Gamma^-} + (1-c_3)(1+z)^{3\Gamma^+}\Big),}
\een
where $\bar\rho$ and $c_3$ are constants of integration.\\
With the expressions (\ref{03}), (\ref{54}), (\ref{55}) and (\ref{56}) we can obtain all the explicit functions of the redshift describing the total energy density $\rho^{(\mathcal{Q}_2)}_t(z)$, the  dark single energy densities $\rho^{(\mathcal{Q}_2)}_1(z)$ and $\rho^{(\mathcal{Q}_2)}_2(z)$, the ratio $r^{(\mathcal{Q}_2)}(z)=\rho^{(\mathcal{Q}_2)}_1(z)/\rho^{(\mathcal{Q}_2)}_2(z)$, the $\mathcal{Q}_2$ interaction, the deceleration parameter $q^{(\mathcal{Q}_2)}$, all the effective EoS $\omega_{1eff}^{(\mathcal{Q}_2)}$, $\omega_{2eff}^{(\mathcal{Q}_2)}$ and $\omega_{t}^{(\mathcal{Q}_2)}$ and the density parameter of the dark energy $\Omega_2^{(\mathcal{Q}_2)}$:

\ben
\n{57}
\textstyle{\frac{\rho^{(\mathcal{Q}_2)}_t(z)}{\rho_0}}=\textstyle{\Big[b_mx^{\gamma_m}+(1-b_m)\Big(c_3 x^{\Gamma^-}+(1-c_3) x^{\Gamma^+}\Big)\Big]},\ \ \ 
\een

\ben
\n{58}
\textstyle{\frac{\rho^{(\mathcal{Q}_2)}_1(z)}{\rho_0}}= \textstyle{\frac{(1-b_m)\Big(c_3 (\Gamma^- -\gamma_2)x^{\Gamma^-}+(1-c_3)(\Gamma^+-\gamma_2) x^{\Gamma^+}\Big)}{(\gamma_1-\gamma_2)}}, 
\een

\ben
\n{59}
\textstyle{\frac{\rho^{(\mathcal{Q}_2)}_2(z)}{\rho_0}}= \textstyle{\frac{(1-b_m)\Big(c_3 (\gamma_1-\Gamma^-)x^{\Gamma^-}+(1-c_3)(\gamma_1-\Gamma^+) x^{\Gamma^+}\Big)}{(\gamma_1-\gamma_2)}}, 
\een

\ben
\n{60}
\textstyle{r^{(\mathcal{Q}_2)}(z)}=\textstyle{\frac{\Big(c_3 (\Gamma^- -\gamma_2)x^{\Gamma^-}+(1-c_3)(\Gamma^+-\gamma_2) x^{\Gamma^+}\Big)}{\Big(c_3 (\gamma_1-\Gamma^-)x^{\Gamma^-}+(1-c_3)(\gamma_1-\Gamma^+) x^{\Gamma^+}\Big)}},
\een

\begin{multline}
\n{61}
\textstyle{\frac{\mathcal{Q}_2(z)}{\rho_0(1-b_m)}}= \textstyle{\frac{\sigma \Big(c_3(k-\Gamma^-) x^{\Gamma^-}+(1-c_3)(k-\Gamma^+) x^{\Gamma^+}\Big)}{(\gamma_1-\gamma_2)}},
\end{multline}

\be
\n{62}
\textstyle{1+q^{(\mathcal{Q}_2)}(z)}=\textstyle{\frac{3}{2} \frac{b_m\gamma_mx^{\gamma_m}+(1-b_m)\Big(c_3\Gamma^-x^{\Gamma^-}+(1-c_3)\Gamma^+ x^{\Gamma^+}\Big)}{b_mx^{\gamma_m}+(1-b_m)\Big(c_3x^{\Gamma^-}+(1-c_3) x^{\Gamma^+}\Big)},}\ \ 
\ee

\be
\n{63}
\textstyle{\Omega_2^{(\mathcal{Q}_2)}(z)}=\textstyle{\frac{(1-b_m)}{(\gamma_1-\gamma_2)}\frac{\Big(c_3 (\gamma_1-\Gamma^-) x^{\Gamma^-} + (1-c_3) (\gamma_1-\Gamma^+) x^{\Gamma^+}\Big)}{\Big(b_m\gamma_mx^{\gamma_m}+(1-b_m)[c_3x^{\Gamma^-}+(1-c_3) x^{\Gamma^+}]\Big)}},
\ee

\be
\n{64}
\textstyle{\omega_{1eff}^{(\mathcal{Q}_2)}(z)=\omega_1}\textstyle{+\sigma \frac{ \Big(c_3 (k - \Gamma^- )x^{\Gamma^-}+(1-c_3)(k - \Gamma^+) x^{\Gamma^+}\Big)}{\Big(c_3 (\Gamma^- -\gamma_2)x^{\Gamma^-}+(1-c_3)(\Gamma^+-\gamma_2) x^{\Gamma^+}\Big)},}
\ee

\be
\n{65}
\textstyle{\omega_{2eff}^{(\mathcal{Q}_2)}(z)}=\textstyle{\omega_2- \sigma \frac{ \Big(c_3 (k - \Gamma^- )x^{\Gamma^-}+(1-c_3)(k - \Gamma^+) x^{\Gamma^+}\Big)}{\Big(c_3 (\gamma_1 - \Gamma^- )x^{\Gamma^-}+(1-c_3)(\gamma_1 - \Gamma^+) x^{\Gamma^+}\Big)},}
\ee

\be
\n{66}
\textstyle{\omega_{t}^{(\mathcal{Q}_2)}(z)}=\textstyle{\frac{ b_m \omega_mx^{\gamma_m}+(1-b_m)\Big(c_3\omega_- x^{\Gamma^-}+(1-c_3)\omega_+ x^{\Gamma^+}\Big)}{b_mx^{\gamma_m}+(1-b_m)\Big(c_3 x^{\Gamma^-}+(1-c_3) x^{\Gamma^+}\Big)},}\ \ \ \ 
\ee
\no with $e^{-\eta}=(1+z)^3\equiv x$, $\omega_m=\gamma_m-1$,  $\omega_{\pm}=\Gamma^{\pm}-1$ and $k\equiv (\gamma_1+\alpha\gamma_2)/(1+\alpha)$.

%QsunyueFig1
\begin{figure}
\vskip -0.6cm
\centering 
\begin{minipage}[t][10cm][b]{0,45\textwidth}
\includegraphics[height=30cm,width=19cm]{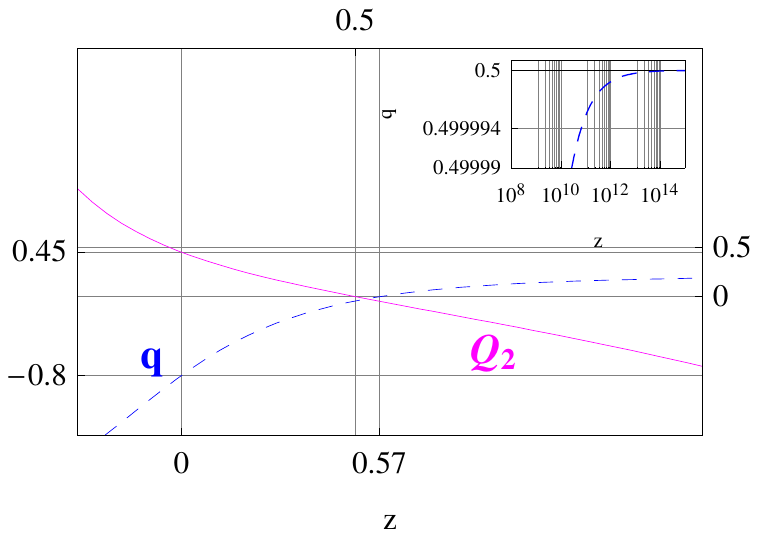}
\caption{\scriptsize{Evolution of interaction $\mathcal{Q}_2$ (in units of $3H_0^2$) and its corresponding decelerating parameter $ q $. The transition to an accelerating universe is checked at $z_{acc} = 0.57$ while the interaction changes sign at $z_{\mathcal{Q}_2} = 0.5$. At early times $q$ tends to 0.5 as in the $\Lambda$CDM but its current value is  $q = - 0.8$, that is, much faster than in that model. The current magnitude of the interaction is $1.35H_0^2$, and is transferring energy from DM to DE.}}
\label{Fig:.QsunyueFig1}
\end{minipage}
\end{figure}

\begin{figure}
\vskip -0.8cm
\centering 
\begin{minipage}[t][10cm][b]{0,46\textwidth}
\includegraphics[height=25cm,width=13cm]{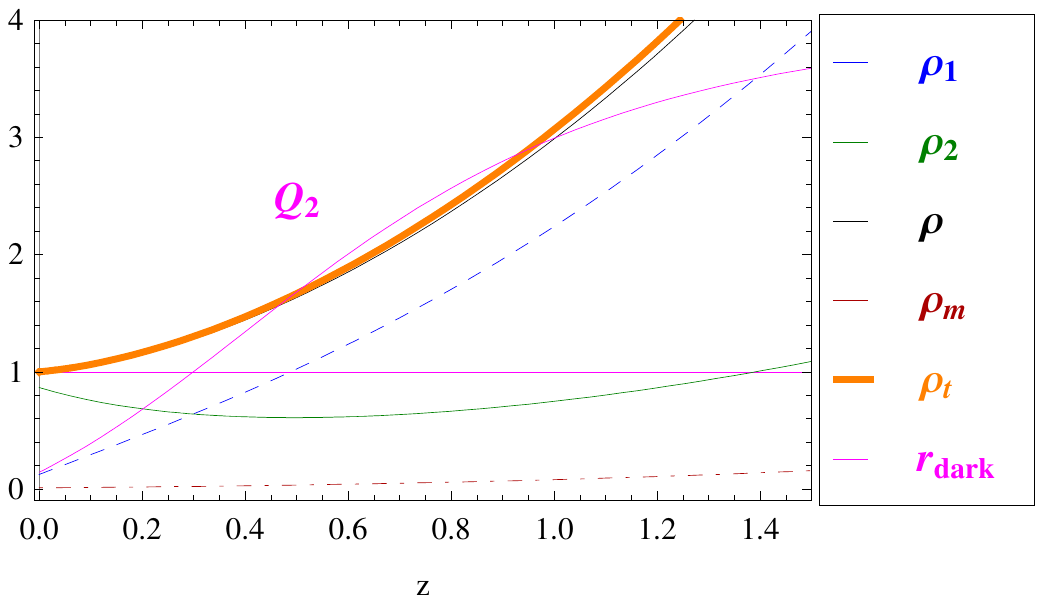}
\caption{\scriptsize{ Evolution of all energy densities (in units of $3H_0^2$), in the best adjusted model with the interaction $\mathcal{Q}_2$. The total energy density is mostly composed by CDM to about $z = 0.3$ where the dark ratio equals the unit. The energy density of DE is a decreasing function of time until $z_{\mathcal{Q}}=0.5$ where the interaction reverses its sign and DE increases at the expense of the CDM.}}
\label{Fig:.QsunyueFig2}
\end{minipage}
\end{figure}

\begin{figure}
\vskip -0.4cm
\centering 
\begin{minipage}[t][10cm][b]{0,46\textwidth}
\includegraphics[height=28cm,width=18cm]{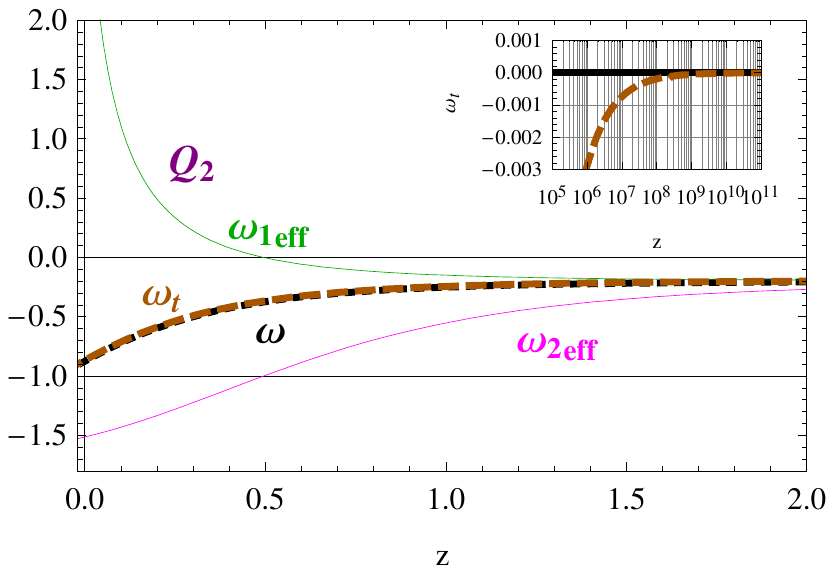}
\caption{\scriptsize{Evolution of all effective EoS in the best fit model with the interaction $\mathcal{Q}_2$. The equivalent fluid behaves as dust at early times, as it can be seen in the inset, and its actual value is  $\omega_{t}(0)=-0.9$. The effective EoS of dark energy has the actual value $\omega_{2eff}(0)=-1.5$, and crossing the PDL at redshift very similar to the transition $z_{acc}=0.57$.}}
\label{Fig:.QsunyueFig3}
\end{minipage}
\end{figure}

\begin{figure}
\vskip 1.3cm
\centering 
\begin{minipage}[t][10cm][b]{0,46\textwidth}
\includegraphics[height=20cm,width=13cm]{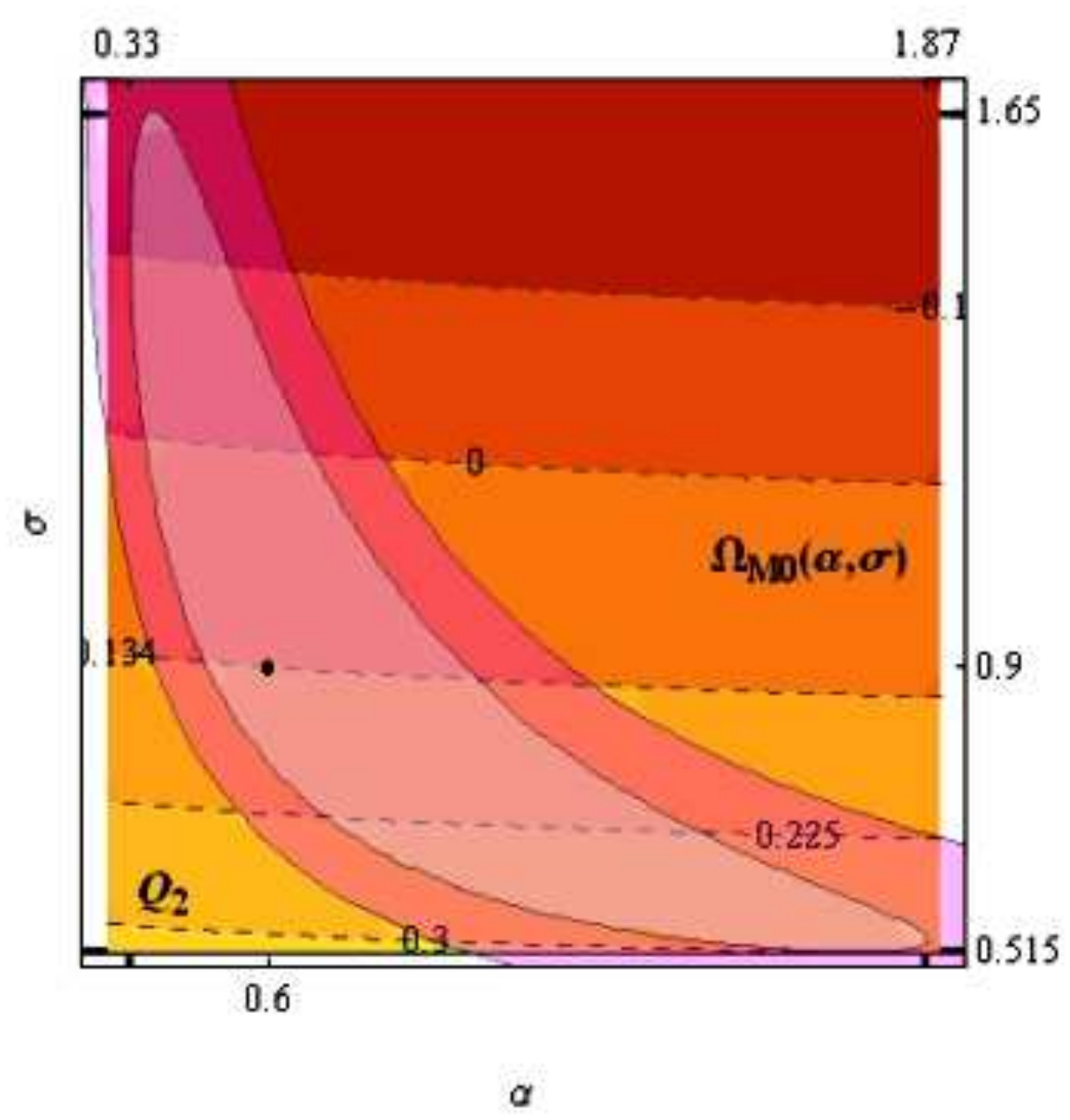}
\caption{\scriptsize{Variation of the sum $\Omega_{M0}$ of the two actual density parameters of matter (Dust plus CDM) in the case $\mathcal{Q}_2$ due to variations at $1\sigma$ CL of the coupling parameters. The figure is obtained overlapping $1\sigma$ confidence region in the parameter space $\alpha$ vs. $\sigma$  onto the contours of the sum $\Omega_{M0}$. From  there we can infer that the current values of total matter content correspond to the range $[0,0.3]$ while the best fit is $\Omega_{M0}=0.134$.}}
\label{Fig:.QsunyueFig4}
\end{minipage}
\end{figure}

\begin{figure}
\vskip 0.3cm
\centering 
\begin{minipage}[t][10cm][b]{0,48\textwidth}
\includegraphics[height=20cm,width=15cm]{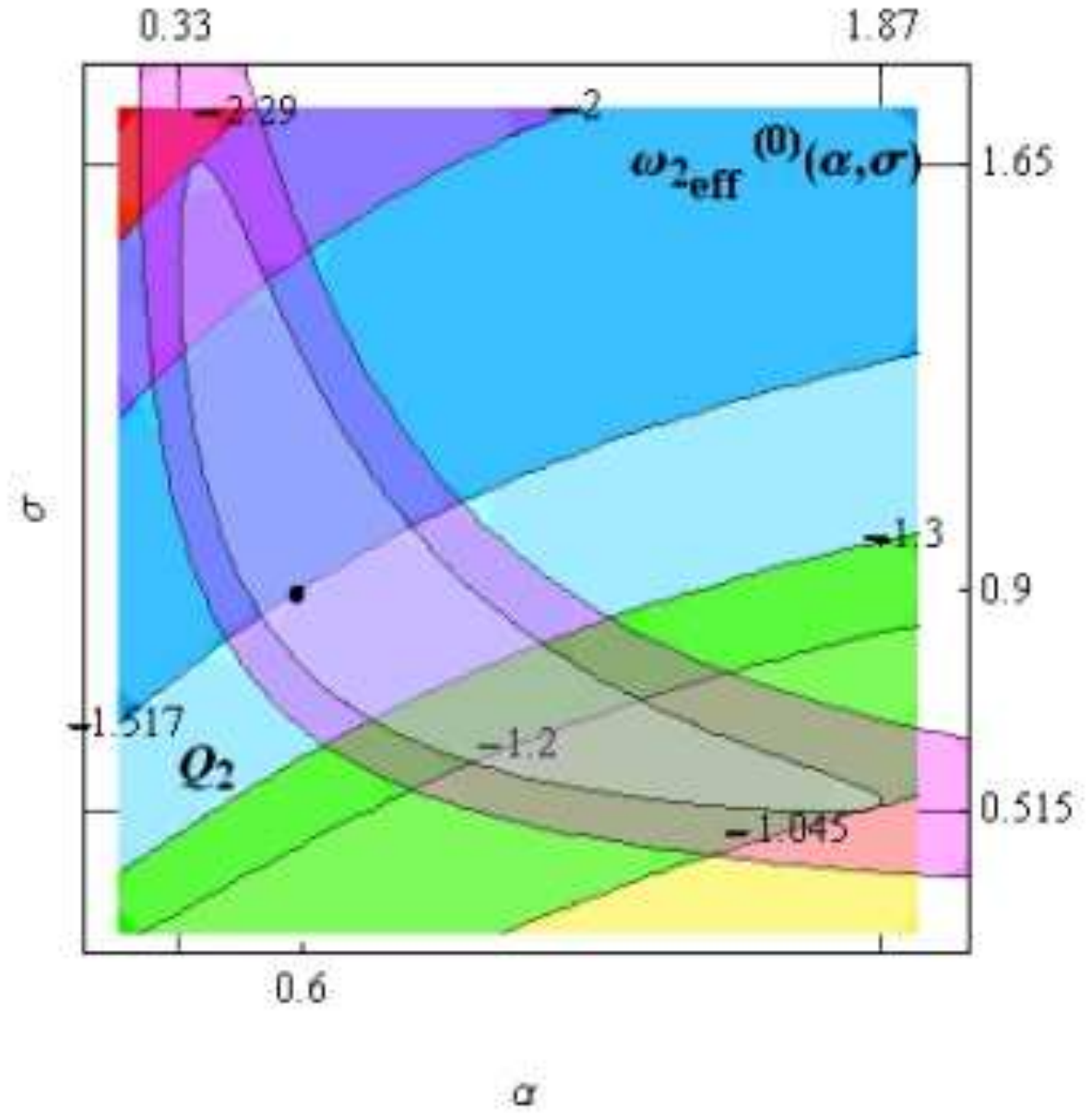}
\caption{\scriptsize{Variation of the actual effective EoS $\omega_{2eff}$  in the case $\mathcal{Q}_2$ due to variations at $1\sigma$ CL of the coupling parameters. The figure is obtained overlapping  $1\sigma$ confidence region in the parameter space $\alpha$ vs. $\sigma$  onto the contours of $\omega_{2eff}(\alpha,\sigma)$. From the contours  we can infer that  the current values correspond to the range $[-2.29,-1.045]$ while the best fit is $\omega_{2eff}=-1.517$.}}
\label{Fig:.QsunyueFig5}
\end{minipage}
\end{figure}

\begin{figure}
\vskip 2.8cm
\centering 
\begin{minipage}[t][10cm][b]{0,48\textwidth}
\includegraphics[height=18cm,width=10cm]{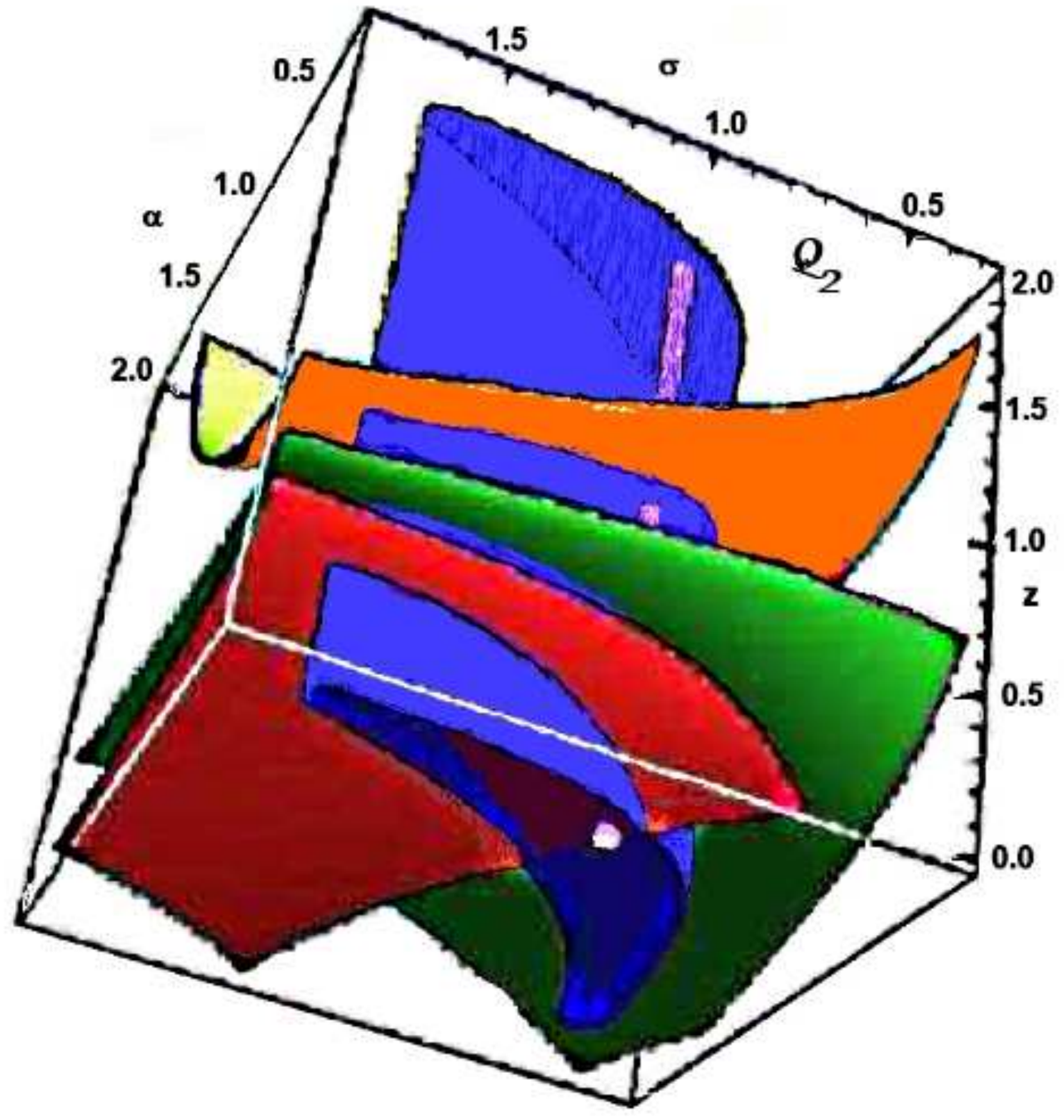}
\caption{\scriptsize{Sample of evolution of $\omega_{2eff}(z)$ at $1\sigma$ CL in the $\alpha$ vs. $\sigma$ parameter space (blue region). In the best fit model for $\mathcal{Q}_2$ (solid pink line), $\omega_{2eff}(z)$ crosses $\omega_{2eff}=-1.5$ (red plane) just before the actual time and at $1\sigma$ CL, it crosses the phantom divide line (green plane) between $z=1$ and $z=0.4$. The yellow plane corresponds to contour  $\omega_{2eff}=-0.5$.}}
\label{Fig:.QsunyueFig6}
\end{minipage}
\end{figure}

\begin{figure}
\vskip -0.2cm
\centering 
\begin{minipage}[t][10cm][b]{0,48\textwidth}
\includegraphics[height=20cm,width=12cm]{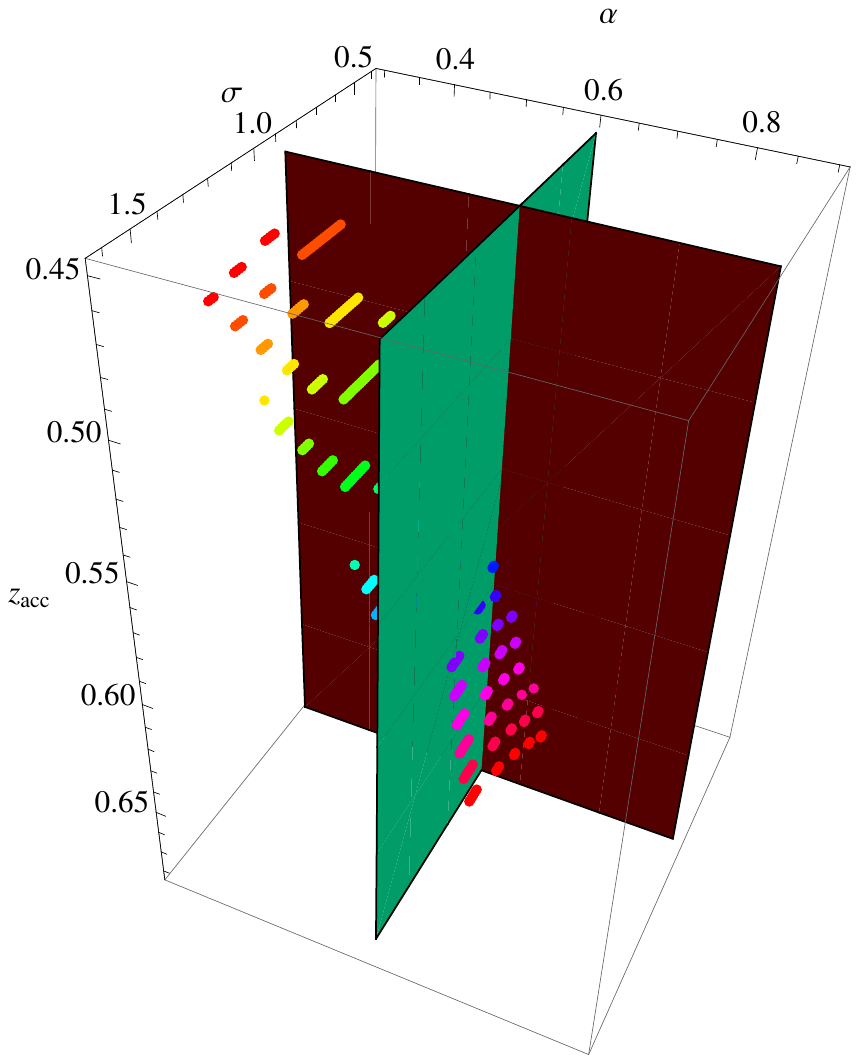}
\caption{\scriptsize{Variation of the redshift of transition to accelerated regime as a function of the interaction parameters $\alpha$ and $\sigma$. At $1\sigma$ CL in the $ \alpha$ vs. $\sigma$ parameter space, $z_{acc}$ is a strongly increasing function of $\alpha$ and a growing one respect to $\sigma$ but with  little sensitivity. This behavior is reflected on the constant planes $ \alpha = 0.6$ (cyan) and $\sigma = 0.9$ (brown).}}
\label{Fig:.QsunyueFig7}
\end{minipage}
\end{figure}

In this case, where $\Theta_{\mathcal{Q}_2}=(H_0,b_m,c_3,\gamma_m,\gamma_1,\gamma_2,\alpha,\sigma)$ and $H_{th}(\Theta_{\mathcal{Q}_2}; z_i)= \sqrt{\rho_t(z_i)/3}$ is taken from (\ref{57}),  we find $\chi^2_{min}=18.68$, corresponding to  $\chi^2_{dof}=0.8895$, for the best fit values: $H_0=70.64~{\rm km s^{-1}\,Mpc^{-1}}$, $b_m=0.01$, $c_3=0.45$, $\gamma_m=1.00$, $\gamma_1=1.00$, $\gamma_2=0$, $\alpha=0.6$ and $\sigma=0.9$. 
That is, the best adjusted model for $\mathcal{Q}_{2}$, supports an interaction between CDM and cosmological constant $\Lambda $ in presence of non interactive dust. 
In Fig.:\ref{Fig:.QsunyueFig1}, the interaction $\mathcal{Q}_2$ and the corresponding deceleration parameter $q^{(\mathcal{Q}_2)}$ are drawn as functions of the redshift. In this model there is strong energy transfer from DE to the particles of CDM at early times and equally strong is the energy transfer from the CDM decaying into DE in the distant future.  Instead in recent history, the transference  is relatively smooth and  CDM particles creation ceases from $z_{\mathcal{Q}}=0.5$ when its effects are reversed. For $z >>1$, $q$ tends asymptotically to a  value $q_e=0.5$ mimicking $\Lambda$CDM at early times and the same is true for the current time when  $q_0=-0.8$. This value coincides with the correspondent to the model $\Lambda$CDM, calculated with the parameter of density of both, interactive plus not interactive dust $\Omega_{M0}$ and the transition redshift $z_{acc}=0.57$ is consistent with the values consigned in literature \cite{Lima:2012bx}, \cite{Cunha:2008mt},\cite{Liu:2012me}, (See Table \ref{tab:tabla2}). 
Fig.:\ref{Fig:.QsunyueFig2}. shows the evolution of all  energy densities $\rho_1$, $\rho_2$, $\rho_m$, $\rho$ and $\rho_t$ for the best fit model and also of the ratio $r=\rho_1/\rho_2$ for this  interaction. It can be seen that the relation between the magnitudes of  dark densities  is $r=1$  around $z\sim 0.3$, (later than all the previous cases) and that the total energy is mainly composed of energy of the interacting dark sector. Notably, the dark energy density decreases monotonously until $z=0.3$ when it begins to recover, growing up to the current time reflecting the change of sign of the interaction.
The evolutions of all effective EoS, whose early - time asymptotic values are $\omega_{1eff}= \omega_{2eff}= -0.2$  and $\omega_{t}=0$, are shown in Fig.:\ref{Fig:.QsunyueFig3}. The effective EoS of DE crosses the PDL around $z=0.5$, redshift almost coincident with the redshift of the transition to the accelerated regime, $z_{acc}=0.57$. 
The variations of the actual value of the total matter content $\Omega_{M0}(\alpha,\sigma)$ and of the effective EoS of DE $\omega_{2eff}(\alpha,\sigma)$  due to variations at $1\sigma$ CL of the coupling parameters in the case $\mathcal{Q}_2$ are drawn in Fig.:\ref{Fig:.QsunyueFig4} and Fig.:\ref{Fig:.QsunyueFig5}. In the first of these two graphs, the $1\sigma$ error ellipse for $\alpha$ vs. $\sigma$  parameters space,  superimposed onto contours of sum of the density parameters $\Omega_{M0}(\alpha,\sigma)$, shows that  $\Omega_{M0}=0.134^{+0.166}_{-0.134}$. In the second graph, the same error ellipse indicates 
that the actual $\omega_{2eff}=-1.517^{+0.472}_{-0.773}$.
The Fig.:\ref{Fig:.QsunyueFig6} is a sample of evolution of the effective equation of state of DE where it can be seen that  the best fit model for $\mathcal{Q}_2$ (solid pink line), $\omega_{2eff}(z)$ crosses $\omega_{2eff}=-1.5$ (red plane) just before the actual time and at $1\sigma$ CL in $\alpha$ vs. $\sigma$ parameters space, it crosses the phantom divide line (green plane) between $z=1$ and $z=0.4$. The yellow plane corresponds to contour  $\omega_{2eff}=-0.5$.
Finally, the effect of varying the coupling parameters at $1\sigma$ CL  onto the transition redshift to the accelerated phase  is shown in Fig.:\ref{Fig:.QsunyueFig7}, where it can be seen that $z_{acc}(\alpha,\sigma)$ is a strongly increasing function of $\alpha$ and a growing one respect to $\sigma$ but with  little sensitivity. This behavior is reflected on the constant planes $ \alpha = 0.6$ (cyan) and $\sigma = 0.9$ (brown). Its values belong to the interval $[0.45,0.65]$ when $\alpha \ \epsilon [0.33,0.87]$ and  $\sigma \ \epsilon [0.515,1.65]$.

%QsunyueFig2

%QsunyueFig3

%QsunyueFig4

%QsunyueFig5

%QsunyueFig6

%QsunyueFig7

%%%%%%%%%%%%%%%%%%%%%%%%%%%%%%%%%%%%%%%%%%%%%%%%%%%%%%%%%%%%%%%%%%

\section{Comparisons}

\begin{figure}
\vskip -1.2cm
\centering 
\begin{minipage}[t][10cm][b]{0,48\textwidth}
\includegraphics[height=18cm,width=12cm]{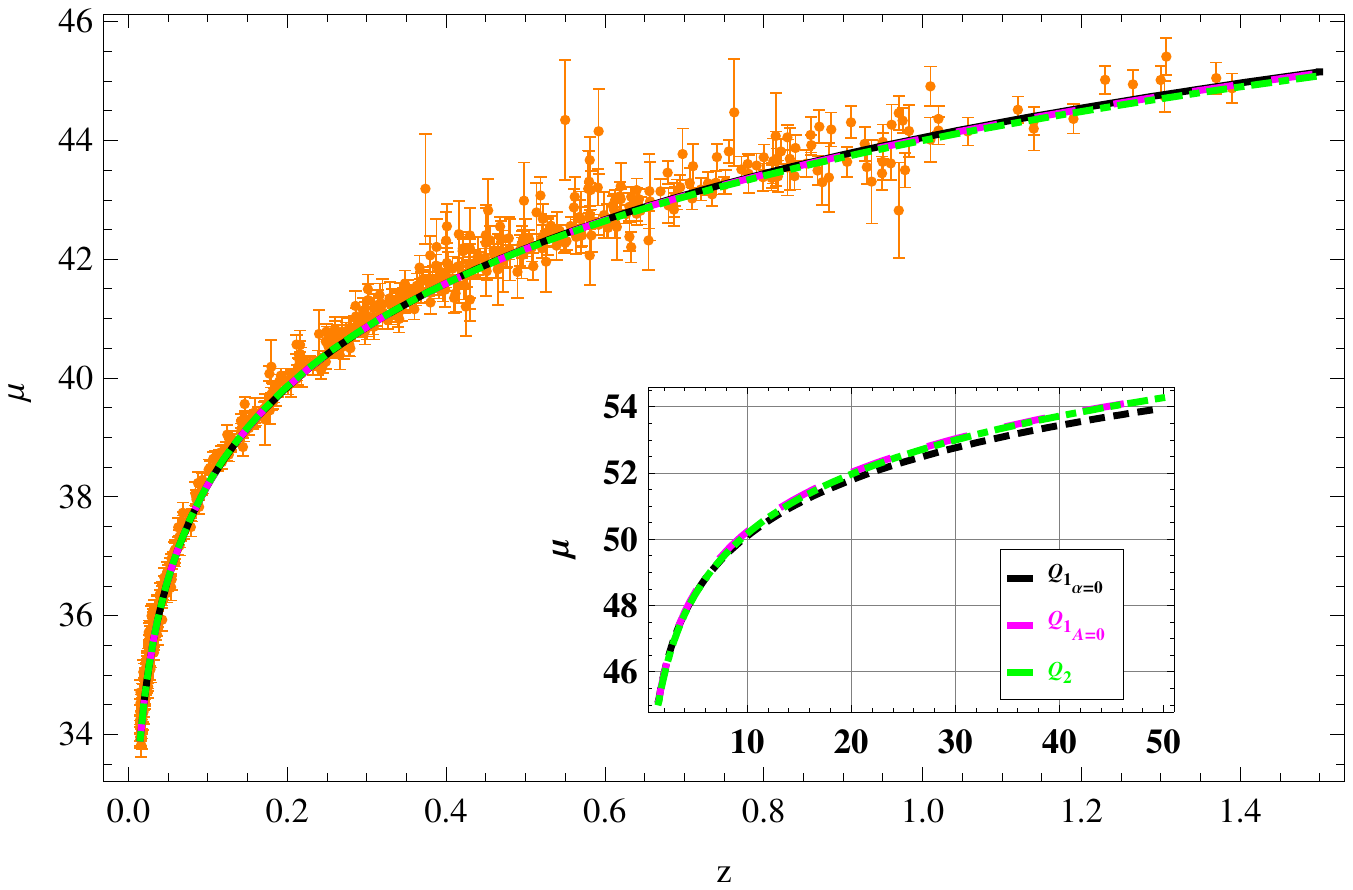}
\caption{\scriptsize{The distance modulus curves for the three models with the better adjusted parameters through SNIA Union2.1 database are compared. Behaviors are identical and fit very well to the experimental data. The differences among the theoretical curves are only seen to higher redshifts  as shown in the inset for the interval between $z=1.5$ and $ z=15$.}}
\label{Fig:.muComparados}
\end{minipage}
\end{figure}

\begin{figure}
\vskip -2.2cm
\centering 
\begin{minipage}[t][11cm][b]{0,48\textwidth}
\includegraphics[height=24cm,width=16cm]{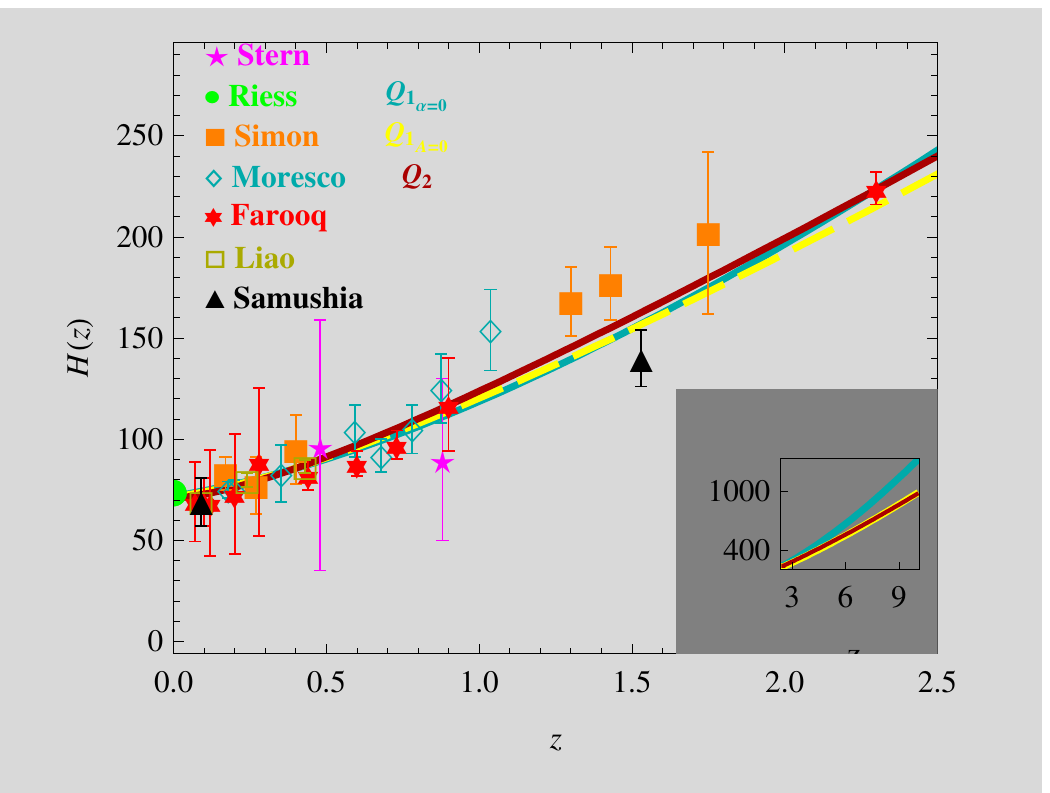}
\caption{\scriptsize{ The curves corresponding to the theoretical Hubble functions of the three models are compared between themselves and with the  Hubble function database. One can see that they are very similar in the range defined by the data, and differ for higher redshifts forming two distinct groups. On one hand the case with  $\mathcal{Q}_{1_{\alpha=0}}$ and on the other   the cases with  $\mathcal{Q}_{1_{A=0}}$ and  $\mathcal{Q}_2$, are the two different sets as it is shown in the inset for the interval between $z=2.5$ and $ z=10$.}}
\label{Fig:.wWeiH}
\end{minipage}
\end{figure}

The curves of distance modulus for the models with the better adjusted parameters are compared in Fig.:\ref{Fig:.muComparados} with the SNIa Union2.1 database \cite{Suzuki:2011hu}  for these three interactions. Their behaviors are identical and fit very well to the experimental data. The difference among the theoretical curves are only seen at higher redshifts  as shown in the inset for the interval between $z=1.5$ and $ z=50$.   This means that the supernova data are not the appropriate tool to choose among these models because Union2.1 upper bound corresponds to a redshift $z=1.39$  and the curves are separated later. 
Better discrimination is obtained on Fig.:\ref{Fig:.wWeiH}  where the curves corresponding to the theoretical Hubble functions of the three models are compared between themselves and with respect to the Hubble data function  \cite{Stern:2009ep,Riess:2009pu,Simon:2004tf,Moresco:2012by,Farooq:2013hq,Liao:2012bg}. There it can be seen that they are very similar in the range defined by the data but later, they differ for higher redshifts forming two distinct groups. On one hand, the case with  $\mathcal{Q}_{1_{\alpha=0}}$ and on the other, the cases with  $\mathcal{Q}_{1_{A=0}}$ and  $\mathcal{Q}_2$, are two different sets as it is shown in the inset for the interval between $z=2.5$ and $ z=10$. These two forms of comparison do not turn out to be satisfactory to incline our preferences in favor of anyone of the studied interactions.

%figura 22

%figura 23

%figura 24
\begin{figure}
\vskip -2.2cm
\centering 
\begin{minipage}[t][11cm][b]{0,46\textwidth}
\includegraphics[height=24cm,width=18cm]{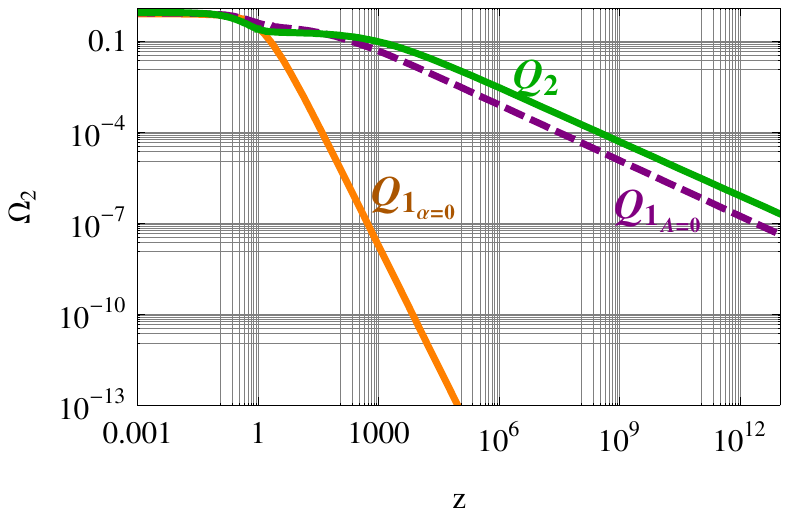}
\caption{\scriptsize{Comparison of  the evolution of the density parameter $\Omega^{\mathcal{Q}_i}_2$ for the dark energy in the three models studied. There are two types of behavior. For $\mathcal{Q}_{1_{\alpha=0}}$ (solid orange curve) the slope is  very steep  and seems be able to accommodate any restrictions, but it is even more negligible than the cosmological constant. In the other two cases, for $\mathcal{Q}_{1_{A=0}}$ (dashed purple curve) and $\mathcal{Q}_{2}$ (solid green curve), the density parameters $ \Omega_2 $ decrease more slowly and may be consistent with some bounds obtained from  non interacting models, retaining a reasonable amount of energy during the time of the BBN ($\Omega^{\mathcal{Q}_2}_2\sim 10^{-4}$ and $\Omega^{\mathcal{Q}_{1_{A=0}}}_2\sim 10^{-5}$). The dashed curve is derived from an interaction that maintains the term proportional to $\dot\rho_2$. }}
\label{Fig:.EDEcomparados}
\end{minipage}
\end{figure}

%EDE

To further explore the suitability of each interaction, we analyze the behavior of dark energy at early times (EDE). Many authors have claimed the existence of a non-negligible dark energy at early times. That is, the dark energy could not fade away to the $ \sim 10^{-9 }$ fraction of the energy density at CMB recombination that is predicted by the cosmological constant. For example, Wetterich proposes the modeling of the vacuum energy by a phenomenological parameterization of a quintessence and draws as conclusion that EDE should necessarily  exist, under certain circumstances (parameter $b>0$, see \cite{Wetterich:2004pv}). Moreover, some descriptions of this energy, coming from the field of particle physics, involve scalar fields whose densities correspond to a constant fraction of the energy density of the dominant component \cite{Ferreira:1997hj}. In addition, Big bang nucleosynthesis (BBN) ($z\sim 10^8 - 10^9$) can help to determine whether or not there is EDE, constraining the allowable amount of it that does not interfere with the process of nucleosynthesis. The presence of vacuum energy can be seen also, via the expansion speed up provided by this additional energy density. The vacuum energy acts essentially as an extra neutrino species and the bound on the number of species determines a maximum value of the vacuum energy during the baryogenesis era $\Omega_2(1 MeV) < 0.21$ \cite{Cyburt:2004yc}. Fixing the effective number of relativistic neutrinos to the standard value $N^{\nu}_{eff}=3.046$, Calabrese et al.\cite{Calabrese:2011hg}, improve the bound yielding $\Omega_{EDE} < 0.043$ in case of a relativistic EDE and $\Omega_{EDE} < 0.024$ for a quintessence EDE, at BBN epoch. 
Also, Sievers et al. \cite{Sievers:2013ica} say that the small-scale damping seen in the Atacama Cosmology Telescope (ACT) data can be interpreted as arising from an amount, non-negligible, of dark energy at decoupling and their  analysis of the ACT data in combination with WMAP7 data and the  ACT detection measurement, leads to the upper bound $\Omega_{EDE} < 0.025$  (WMAP7 + ACT + ACTDfl, at $95\%$ CL),  with the bound on the equation of state  $\omega_0 < -0.45$. The same model used by Sievers, was previously constrained from Reichardt et al. \cite{Reichardt:2011fv} who reports an upper limit of $\Omega_{EDE} < 0.018$ at $95\%$ CL from CMB only data, combining WMAP and SPT. 
The first cosmological results of the Planck Collaboration \cite{Ade:2013zuv} refer to the two main effects of EDE: the reduction of structure growth in the period after last scattering and the change of position and height of the peaks in the CMB spectrum. Note that the possibility of a dark energy late, after the last scattering surface, is dependent on exactly what supplementary data are used in conjunction with the CMB data. Using the direct measurement of $H_0$, or the SNLS SNe sample, together with Planck shows preferences for dynamical dark energy at about $2\sigma$ CL reflecting the tensions between these data sets and Planck in the $\Lambda$CDM model. 
In contrast, the BAO measurements together with Planck give tight constraints which are consistent with a cosmological constant. Interestingly, the presence or absence of dark energy at the epoch of last scattering ($ z=1100$) is the dominant effect on the CMB anisotropies and hence the constraints are insensitive to the addition of low redshift supplementary data such as BAO. The most precise bounds on EDE arise from the analysis of CMB anisotropies \cite{Doran:2001rw,Caldwell:2003vp,Calabrese:2010uf,Reichardt:2011fv,Sievers:2013ica,Hou:2012xq,Pettorino:2013ia}. Using Planck$+$WP$+$highL, (WP stands for a WMAP polarization low multipole
likelihood  at $\it{l} \leq 23$ and highL is the high-resolution CMB data), $\Omega_{EDE} < 0.009 $ at $95\%$ CL and $\Omega_{EDE} < 0.010 $ for Planck$+$WP only \cite{Bennett:2012zja}.
 These bounds  improve the recent ones of \cite{Hou:2012xq}, who give 
$\Omega_{EDE} < 0.013$ at $95\%$ CL, and  \cite{Sievers:2013ica}, who find $\Omega_{EDE} < 0.025 $ at $95\%$ CL.  
Calabrese et al. \cite{Calabrese:2011nf} discuss present and future cosmological constraints on variations of the fine structure constant $\alpha$ induced by an EDE component having linear coupling to electromagnetism. They found no variation of the fine structure constant at recombination respect to the present-day value, constraining EDE to $\Omega_{EDE} < 0.060$ at $95\%$ CL. Hollenstein et al. \cite{Hollenstein:2009ph} constraint EDE from CMB lensing and weak lensing tomography finding a best fit 
$\Omega_{EDE} \sim 0.03$ with different $1\sigma$ errors according  to the different experiments and the combinations of Euclid with Planck and CMBPol.
In summary, we have a variety of restrictions on the EDE, resulting from its possible action on cosmological structure formation, variation of the fine structure constant, growth of fluctuations in dark matter and primordial nucleosynthesis, but they all refer to models without interaction in the dark sector and there are also considerations about the temporary location (or redshift) at which they must be applied \cite{Pettorino:2013ia,Cahill:2013kda}. The Fig.:\ref{Fig:.EDEcomparados}  shows the evolution of the curves corresponding to density parameter of dark energy  for the three models considered, for which clearly, there are two types of behavior. For the case with  interaction $\mathcal{Q}_{1_{\alpha=0}}$ (solid orange curve) the slope is very steep and able to accommodate any restrictions, but it is even more negligible than the cosmological constant.  In the other two cases, with interactions $\mathcal{Q}_{1_{A=0}}$ (dashed purple curve) and $\mathcal{Q}_{2}$ (solid green curve), the density parameters $ \Omega_2 $ decrease more slowly and may be consistent with some bounds obtained from  non interacting models.
It must be highlighted the fact that the behaviors drawn in Fig.:\ref{Fig:.EDEcomparados} correspond to models with interactions described in the dark sector plus the inclusion of a non-interactive material component, so that before considering whether or not they satisfy the aforementioned constraints, they should be considered as examples of the possible values of EDE in the class of interactive scenarios.

%%%%%%%%%%%%%%%%%%%%%%%%%%%%%%%%%%%%%%
%%Crisis de la edad

\begin{figure}
\vskip -2.2cm
\centering 
\begin{minipage}[t][11cm][b]{0,46\textwidth}
\includegraphics[height=24cm,width=16cm]{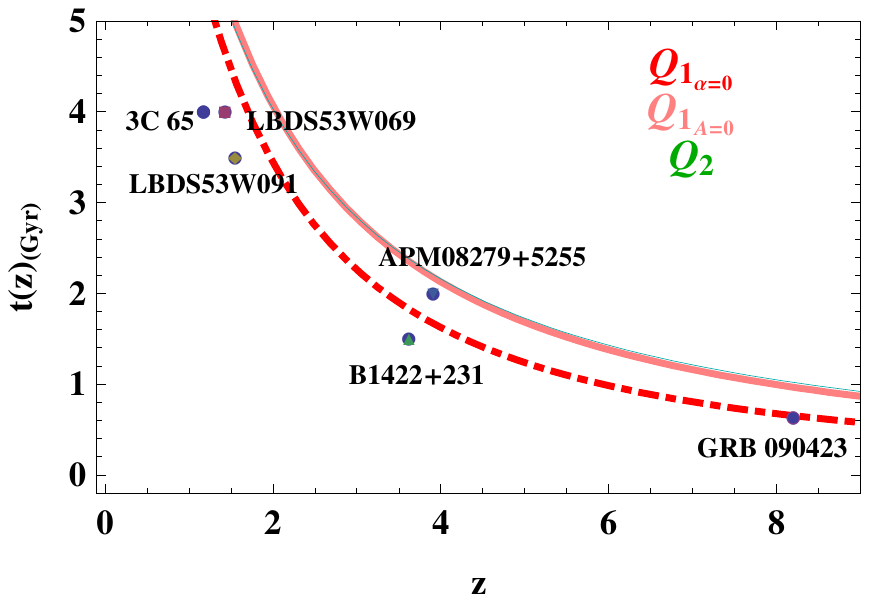}
\caption{\scriptsize{Comparison of the curves of time-redshift relation with best-fit parameters for the three models studied, among themselves and with the old high redshift objects (OHROs).The milestones considered are  old ratio galaxy 3C 65, old galaxy LBDS $53W069$, old galaxy LBDS $53W091$, high redshift quasar B$1422+231$, old quasar APM $08279+5255$ and the most distant GRB $090423$. It is observed that the crisis of cosmological age is solved for the cases with interactions $\mathcal{Q}_{1_{A=0}}$ and $\mathcal{Q}_2$ but the objects APM0$8279 + 5255$  and GRB $090423$ are still  pebbles in its shoes for $\mathcal{Q}_{1_{\alpha=0}}$.}}
\label{Fig:.Crisis}
\end{minipage}
\end{figure}

Now we are going to pay attention to the so-called ``crisis of the age" in cosmology, which refers to the problem of having a universe younger than its constituents, and to its use to constrain cosmological models. This issue appears to have been noticed for the first time by Komossa et al.  \cite{Komossa:2002cn}, in the context of a matter dominated FRW universe and the use of abundance ratio of $Fe / O$ as cosmic clock. Moreover, the X-ray results of Hasinger et al.  on the broad absorption line (BAL) quasar APM $08279+5255$ at redshift $z=3.91$, that is, the high ratio $Fe / O \sim 3$ (equivalent to a 3 Gyr old object), imply that SN Ia are involved \cite{Hasinger:2002wg}, because of the inability to obtain these ratios in SN II. The introduction of dark energy alone does not remove the problem with this quasar and in \cite{Wang:2008te} it has been shown that the
interaction between dark matter and dark energy needs to be taken into account.  Also, in the context of the holographic models  \cite{Cui:2010dr}, it has been achieved that to alleviate the problem, the interaction used must involve not only the dark  holographic energy but also spatial curvature. The new agegraphic dark energy (NADE) models that include interactions between DE and DM (for example $\mathcal{Q}_2$ with $\alpha=1$ and $\sigma=2b$) as in \cite{Li:2010ak} have been able to accommodate this annoying object that has eluded be tamed in a huge variety of proposals \cite{Alcaniz:1999kr,Friaca:2005ba,Alcaniz:2003fy,Yang:2009ae,Wang:2010su,Dantas:2006dy,Capozziello:2007gr,Movahed:2007cs,Movahed:2007ie,Movahed:2007ps,Pires:2006rd,Rahvar:2006tm,Tong:2009mu,Wei:2007ig,Zhang:2007ps,Duran:2010ky,Forte:2012ww}.  Regardless of curvature, it seems still possible to alleviate the problem caused by the OHRO at $z=3.91$, \cite{Duran:2010hi}, but the age problem becomes serious when we consider  the age of the universe at high redshift and it is a good test to be applied to the models proposed. To that end, we use some old high redshift objects (OHROs) discovered, for instance, the 3.5 Gyr old galaxy LBDS 53W091 \cite{Dunlop:1996mp,Spinrad:1997md} at redshift $z = 1.55$,  the 4.0 Gyr old galaxy LBDS 53W069 \cite{Dunlop:1998tm} at redshift $z = 1.43$, the 4.0 Gyr old radio galaxy 3C 65 at $z = 1.175$ \cite{StocktonKelloggRidgway}\cite{LacyRawlingsEalesDunlop}, the high redshift quasar B1422+231 at $z = 3.62$ \cite{Yoshii:1998bw} whose best-fit age is 1.5 Gyr, besides of the aforementioned old quasar APM 08279+5255 at $z = 3.91$ with 3.0 Gyr. To assure the robustness of our analysis, we use the 0.63 Gyr gamma-ray burst GRB 090423, at $z = 8.2 $ \cite{Tanvir:2009zz,Salvaterra:2009ey} detected by the Burst Alert Telescope (BAT) on the Swift satellite 18 on 23 April 2009. This is well beyond the redshift of the most distant spectroscopically confirmed galaxy ($z=6.96$) and quasar ($z=6.43$). 
These OHRO's are not the only ones available to qualify our interactions. In a recent work \cite{Balestra:2013gra}, the VIMOS/VLT observations providing a spectroscopic confirmation at $z=6.110$ for a galaxy quintuply imaged by the Frontier Fields cluster RXC $J2248.7-4431$. These results, together with those recently presented by \cite{Monna:2013eia}, suggest that this magnified, distant galaxy is a young with age less than 0.3 Gyr.  In the future, through campaigns such as the Hubble Ultra Deep Field \cite{Ellis:2012bh} we could obtain confirmation of the existence of other sources  at redshift 8.5 to 12 whose estimated ages allow us to improve this comparison.
In Fig.:\ref{Fig:.Crisis} the curves of time-redshift relation with best-fit parameters for the three models studied, are compared among themselves and with the old high redshift objects (OHROs). The milestones considered are  old ratio galaxy 3C 65, old galaxy LBDS $53W069$, old galaxy LBDS $53W091$, high redshift quasar B$1422+231$, old quasar APM $08279+5255$ and the most distant GRB $090423$. It is observed that the crisis of cosmological age is solved for models with interactions $\mathcal{Q}_{1_{A=0}}$ and $\mathcal{Q}_2$ but not for those with $\mathcal{Q}_{1_{\alpha=0}}$ interaction.

%figura 25

%%%Factor de escala

\begin{figure}
\vskip -1.2cm
\centering 
\begin{minipage}[t][11cm][b]{0,46\textwidth}
\includegraphics[height=22cm,width=18cm]{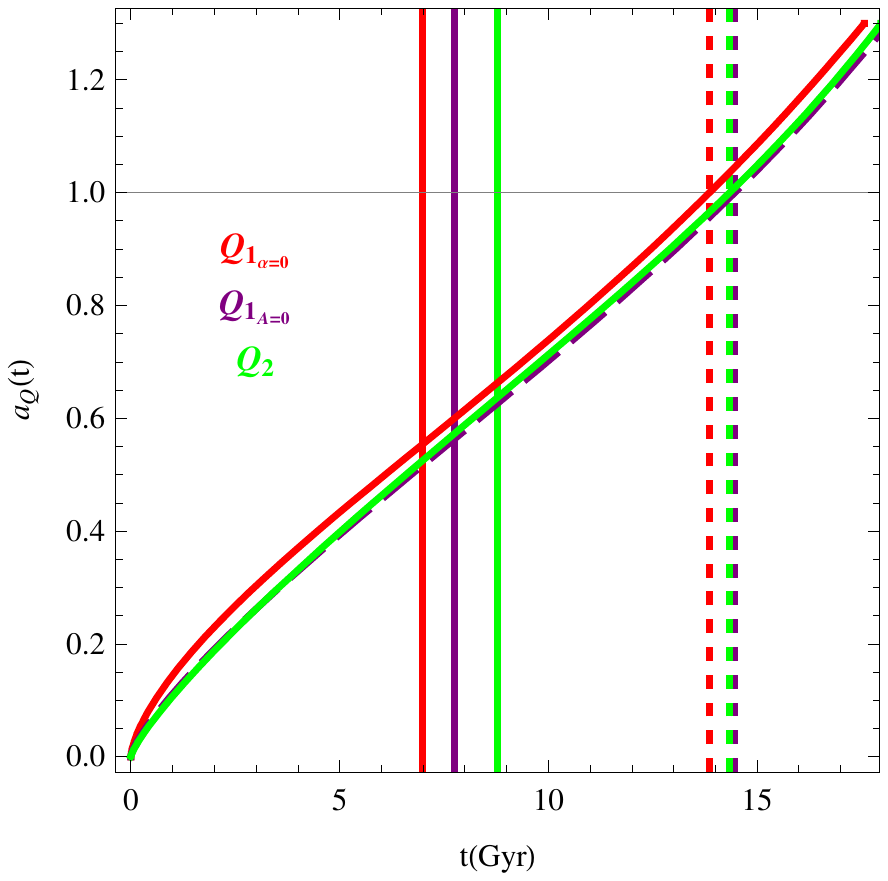}
\caption{\scriptsize{The scale factor curves for each best-fitting model behave very similar although the saddle points, corresponding to the transition to accelerated regime are different (vertical solid lines). The models have different ages of the universe (T in Gyr for which $a (T) = 1$), the youngest for $\mathcal{Q}_{1_{\alpha=0}}$ and the older for  ${\mathcal{Q}_{1_{A=0}}}$ (vertical, thick, dashed lines). }}
\label{Fig:.FactEscala}
\end{minipage}
\end{figure}

The parametric curves for the factor of scale of each  models considered, drawn in Fig.:\ref{Fig:.FactEscala}, allow us to observe  that all the interactions drive  universes not accelerated in the remote past which effect their transitions to accelerated regime at different times (the saddle points marked with solid lines). The youngest universe ($\mathcal{Q}_{1_{\alpha=0}} $ in red) begins its acceleration about 7 Gyr after the Big Bang (BB) and of the other two, the oldest  ($\mathcal{Q}_{1_{A=0}} $ in purple) starts its transition before the other.  The most delayed ($\mathcal{Q}_2$ in green), starts its transition only after 10 Gyr from the BB. The horizontal line $a(t)=1$ marks the present epoch, that is, the age of the universe $T_{\mathcal{Q}_i}$, which for all the models is lower than 15 Gyr. According to the results of Table \ref{tab:tabla3}, the acceleration of the universe may have started between 7 and 9 Gyr after the Big Bang. 
Complementarily we can use the look back time of the beginning of that transition. The look-back time 
$$t^{lb}(z)=\int_0^{z}\frac{dx}{(1+x)\sqrt{\rho_t(x)/3}}$$ is observationally estimated as the difference between the present day age of the universe and the age of a given object at redshift z  \cite{Capozziello:2004jy}\cite{Dalal:2001dt}.  The results correspond to an interval between 5.56 and 6.87 Gyr ago and agree with SN Ia observations, at approximately 6 billion light years ago, realized by two independent groups, led by S. Perlmutter and by B. Schmidt and A. Riess respectively, that revealed the acceleration of the universe (for what they got the Nobel prize in Physics 2011) \cite{Perlmutter:1998np,Riess:1998cb}.

%%%%%%%%%%%%%%%%%%%%%%%%%%%%%%%%%%%%%%%%%%%%%%%%%%
%%%%%%%%%%%%%%%

%%%%%%%%%%%%%%%%%%%%%%%%%%%%%%%%%%%%%%%%%%%%%%%

\section{Conclusions}

We studied three scenarios, with interactions in the dark sector  which are able to change sign along the cosmological evolution, independently of the deceleration parameter and of the ratio of energy densities. We obtained the general solutions for each model, to which we added a material self preserved component. The models were  statistically analyzed using the Hubble function and Tables \ref{tab:tabla1}-\ref{tab:tabla3} summarize the results for the best-fit models for each interaction. The couplings generate DM at the cost of DE, at early times, except in the second one, ${\mathcal{Q}_{1_{A=0}}}$, where the opposite happens. The change of sign for each interaction is  produced inside the interval of redshifts $ [0.48,1.19] $, which is consistent with the results of Cai $\&$ Su, ($0.45<z<0.9$), for cases as ours where $\omega _ {2} =-1$.
Table \ref{tab:tabla1} contains the values of the coupling constants and of the equations of state of interactive and non interactive fluids,  and the current value of the Hubble parameter. Table \ref{tab:tabla2} consigns the current magnitudes of the interactions, of 
density parameters, of the deceleration parameter, the effective equation of state of dark energy  and of the ratio of energy densities of the dark sector. In it, we can see the magnitude of the density parameter of dark energy at the BBN epoch and  also the different redshifts: of  the transitions to the accelerated regime,  of the change of sign by the interactions and those for which the dark ratio is 1, used to qualify the merit of the interaction in alleviating the coincidence problem. Table \ref{tab:tabla3} lists the age of the universe, the time since the BB to the beginnings of the transition to the accelerated regimen and the look back time of these beginnings. Moreover, it describes the $1\sigma$ CL for the coupling parameters $\alpha$ or $\beta$ and $\sigma$ and the ranges of variation for the  transition redshift, for the current effective EoS of dark energy and for the current density parameter of total material component when the coupling constants vary within their $1\sigma$ CL. In the last two columns, Table \ref{tab:tabla3} shows the qualifications of the interactions with respect to  the coincidence problem and their success or failure with respect to the solution of the cosmological age crisis. All models correspond to decelerated universes at early times, which show acceleration at present time and change the direction of energy flows at redshifts consistent with the results of Cai $\&$ Su. The confrontations of the distance modulus or of the Hubble function have not been adequate to test the three models and therefore we have appealed to the analysis of the values of EDE, and to the comparisons of time-redshift relations and parametric curves of the scale factors. 
\begin{widetext}

\begin{table}[htbp]
\begin{center}
\begin{tabular}{|c|c|c|c|c|c|c|c|c|c|c|c|}
\toprule
$\mathcal{Q}_i(\alpha,\beta,\sigma,q_{dark},\rho_1,\rho_2,\rho_2')$ &\multicolumn{8}{|c|}{Best Fit Parameters}&$\chi^2_{min}$&$\chi^2_{dof}$& Best Fit\\
\cline{2-9}
  &\multicolumn{2}{|c|}{\tiny{Coupling constants}}& $\gamma_m $ &$\gamma_ 1 $ & $\gamma_ 2 $&$c_i$& $b_m$&$H_0$&  &   & Model \\
\hline
$\scriptsize{\mathcal{Q}_{1_{\alpha=0}}}=\beta(\sigma+q_{dark})\rho_2$&$\beta = - 0.1\ \ \ $&$\sigma = 0.6$& 1 & 1 & 0 & $c_1=0.05$ & 0.225 & 71.81 & 17.1474 & 0.816543 & \tiny{CDM $ \rightleftharpoons \Lambda$ + Dust} \\
\hline
$\tiny{\mathcal{Q}_{1_{A=0}}}=(\sigma+q_{dark})(\frac{2}{3\gamma_1+2(\sigma-1)} \rho_2'+\beta\rho_2)$&$\beta = - 0.745$&$\sigma= 0.22$& 1 & 1.2 & 0 & $c_2=0.45$ & 0.04 & 71.42 & 17.7573 & 0.845586 &\tiny{WDM $ \rightleftharpoons \Lambda$ + Dust} \\
\hline
$\scriptsize{\mathcal{Q}_2}= \frac{\sigma}{1+\alpha}(\rho_2-\alpha\rho_1)$&$\alpha =\ \ 0.6\ \ \ $&$\sigma= 0.9$& 1 & 1 & 0 & $c_3=0.45$ & 0.01 & 70.64 & 18.6807 & 0.889524 & \tiny{CDM $ \rightleftharpoons \Lambda$ + Dust} \\
\hline
\hline
\end{tabular}
\end{center}
\caption{\scriptsize{Explicit functional form of the interactions and their best fit parameters.  The abbreviation $q_{dark}\equiv (q\rho_t - q_m\rho_m)/(\rho_t-\rho_m)$ with $q_m \equiv (-1+3\gamma_m/2)$, is reduced to the  normal deceleration parameter when we do not consider no interacting fluid. The values of $H_0  $ are given in units of [${\rm kms^{-1}Mpc^{-1}}$].}}
\label{tab:tabla1}
\end{table}
\begin{table}[htbp]
\begin{center}
\begin{tabular}{|c|c|c|c|c|c|c|c|c|c|c|c|c|}
\toprule
$\mathcal{Q}_i$&$\mathcal{Q}_i(z=0)$&$\Omega_1(0)$&$\Omega_2(0)$&$\Omega_m(0)$&$\Omega_{M0}$& $\Omega_{EDE}$& $r_{Dark}$&$q_0$&$\omega^{(0)}_{2eff}$&$z_{acc}$ &$z_{\mathcal{Q}_i}$&$z^{(\mathcal{Q}_i)}$\\
&\scriptsize{$[3H_0^2]$}&&&&\scriptsize{$\big(\Omega_1(0)+\Omega_m(0)\big)$}& $\scriptsize{\big(\Omega_2(10^8)\big)}$ &\scriptsize{$\big(\Omega_{M0}/\Omega_2(0)\big)$} &&&&\scriptsize{$\big(\mathcal{Q}_i(z_{\mathcal{Q}_i})=0\big)$} &$\scriptsize{\big(r(z^{(\mathcal{Q}_i)}) =1\big)}$\\
\hline
$\mathcal{Q}_{1_{\alpha=0}}$& 0.03 & 0.004 & 0.771 & 0.225& 0.229 & 0 &0.297& -0.656&-1.04&0.805&1.19&0.443\\
\hline
$\mathcal{Q}_{1_{A=0}}$&  - 0.20 &0.13 & 0.83 & 0.04 & 0.170 & $5\  10^{-5}$ &0.236&-0.7&-0.763&0.772&0.48&0.7\\
\hline
$\mathcal{Q}_{2}\ \ \ \ $& 0.45 &0.124 & 0.866 & 0.01 & 0.134 & $2\  10^{-4}$  &0.155&-0.8&-1.514&0.570&0.5&0.3\\
\hline
\hline
\end{tabular}
\end{center}
\caption{\scriptsize{Actual magnitude of the interactions, in units of $3 H_0^2$,  current energy parameters, ratio between dark matter and dark energy densities, actual deceleration parameter $q_0$, actual DE effective EoS and redshifts of the transition to the accelerated regime, $z_{acc}$, of the change of sign, $z_{\mathcal{Q}_i}$ and of the unit dark ratio $z^{(\mathcal{Q}_i)}$. The energy parameters of the dark energy at early times, around the BBN epoch, are given for $z=10^8$.}}
\label{tab:tabla2}
\end{table}
\begin{table}[htbp]
\begin{center}
\begin{tabular}{|c|c|c|c|c|c|c|c|c|c|c|}
\toprule
$\mathcal{Q}_i$ &  &$t_{acc}$ & look-back  & $1\sigma$ CL for&$1\sigma$ CL for& $\Delta$& $\Delta$ & $\Delta$&\textit{quality}&solve\\
             &$T_{\mathcal{Q}_i}$&  since BB &time $t_{acc}^{lb}$ & $\alpha$ or $\beta$ & $\sigma$ &$z_{acc}(\alpha,\sigma)$ or &$\omega^{(0)}_{2eff}$& \scriptsize{$\big(\Omega_{M0}\big)$}&\scriptsize{\big($1-t(z^{(\mathcal{Q}_i)})/T$\big)}&crisis\\
  &[Gyr]& [Gyr] &  [Gyr]& & &$z_{acc}(\beta,\sigma)$&&&& of age?\\
\hline
$\mathcal{Q}_{1_{\alpha=0}}$ & $13.86$ &$6.99$ & 6.87 & [-0.37,0.61] & [-3.9,7.3] & [0.74,0.89]&[-1.45,-0.95]&[0,0.3]& 33.13\% & NO \\
\hline
$\mathcal{Q}_{1_{A=0}}$&$14.47$ &$7.76$ & 6.71&[-1.86,-0.54] & [-0.022,0.337]& [0.6,0.9]&[-0.91,0.735]&[0,0.24]& 43.63\%& YES\\
\hline
$\mathcal{Q}_{2}  $&$14.35$ &$8.78$ & 5.57 & [0.33,1.87] & [0.515,1.650] &[0.45,0.65]&[-2.29,-1.045]&[0,0.3]& 24.03\%&YES\\
\hline
\hline
\end{tabular}
\end{center}
\caption{\scriptsize{Ages of universes $T$, ages at the beginning of accelerated regime $t_{acc}$  and look back time of the transition $t_{acc}^{lb}$.  $1\sigma$ CL for the coupling parameters and $\Delta$ variations for the transition redshifts $z_{acc}$, actual dark energy EoS $\omega^{(0)}_{2eff}$ and actual content of matter $\Omega_{M0}$ for the best fit models. $\Delta$ variations correspond to the range of values generated when we allow the coupling parameters vary at $1\sigma$ CL.}}
\label{tab:tabla3}
\end{table}
\end{widetext}

 The most successful of the three interactions has proven to be  $\mathcal{Q}_{1_{A=0}}$ that maintains appropriate residual value of the density parameter of dark energy at early times and get fit all controversial OHRO's for  flat FRW cosmologies, besides having the best factor quality.  All these universes are aged below 15 Gyr and their look back times to the transition to the accelerated regime are around 6 Gyr, consistent with the results of Perlmutter, Schmidt and Riess. Notably, the difference observed in the saddle points of the parametric curves of the factor of scale, could make an interesting way to discriminate between these models if the transition redshift got determined with sufficient accuracy.

%%%%%%%%%%%%%%%%%%%%%%%%%%%%%%%%%%%%%%%%%%%%%%%%%%%%%%%%%%%%%%%%%%

\end{document}